\documentclass[a4paper,11pt]{article}
\usepackage[normalem]{ulem}
\usepackage{siunitx}
\usepackage{comment}
\usepackage{indentfirst}
\usepackage{jcappub} 
\usepackage{amsmath,mathtools}
\usepackage[T1]{fontenc} 
\usepackage{appendix}
\usepackage{xcolor}
\definecolor{pyred}{RGB}{214, 39, 40}
\usepackage{tikz}
\usepackage[compat=1.1.0]{tikz-feynman}
\tikzset{
  square dot/.style={
    draw,
    fill=black,
    shape=rectangle,
    minimum size=3pt,
    inner sep=0pt,
  }
}
\tikzfeynmanset{warn luatex=false}
\usetikzlibrary{shapes,arrows,positioning,automata,backgrounds,calc,er,patterns}
\usetikzlibrary{shapes.geometric}
\makeatletter
\def\tikzfeynman@luatex@required@path{}
\def\tikzfeynman@luatex@required@key{}
\makeatother

\title{\boldmath Breaking parity: the case of the trispectrum from chiral scalar-tensor theories of gravity}

\author[a]{Tommaso Moretti,}
\author[b,c,d]{Nicola Bartolo,}
\author[e]{Alessandro Greco}

\affiliation[a]{Dipartimento di Fisica “E. Pancini”, Università degli Studi di Napoli “Federico II”,\\Compl. Univ. di Monte S. Angelo, Edificio G, Via Cinthia, I-80126, Napoli, Italy}
\affiliation[b]{Dipartimento di Fisica e Astronomia ``G. Galilei'', Universit\`a degli Studi di Padova,\\via Marzolo 8, I-35131, Padova, Italy}
\affiliation[c]{INFN, Sezione di Padova,\\via Marzolo 8, I-35131, Padova, Italy}
\affiliation[d]{INAF, Osservatorio Astronomico di Padova,\\vicolo dell' Osservatorio 5, I-35122 Padova, Italy}
\affiliation[e]{Department of Astronomy, University of Florida,\\211 Bryant Space Science Center, Gainesville, FL 32611, USA}

\emailAdd{tommaso.moretti@unina.it}
\emailAdd{nicola.bartolo@pd.infn.it}
\emailAdd{alessandro.greco@ufl.edu}

\abstract{ 
Recently, possible hints of parity violation have been observed in the connected galaxy four-point correlation function. Although the true origin of the signal from the analysis has been debated, should they have a physical origin, they might point to primordial non-Gaussianity and would be evidence of new physics.

In this work, we examine the single-field slow-roll model of inflation within chiral scalar-tensor theories of modified gravity. These theories, treated here as new Lorentz-breaking theories, extend the Chern-Simons one by including parity-violating operators containing first and second derivatives of the
non-minimally coupled scalar (inflaton) field. This model is capable of imprinting parity-violating signatures in late-time observables, such as the galaxy four-point correlation function. 
We perform an analysis of the graviton-mediated scalar trispectrum of the gauge-invariant curvature perturbation $\zeta(t,\mathbf{x})$ using one of the parity-violating operators of these theories as a case study.  
We estimate that for a set of parameters of the theory it is possible to produce a signal-to-noise ratio for the parity-violating part of the trispectrum of order one without introducing modifications to the single-field slow-roll setup.
Even if the signal found in the analysis turns out to be spurious or if no parity violation is ever detected in the galaxy four-point correlation function, our analysis can be used to constrain the free parameters of these theories.
} 

\makeatletter
\gdef\@fpheader{}
\makeatother

\begin{document}
\maketitle
\flushbottom

\section{Introduction}
\label{Sec:introduction}

The exploration of symmetries holds a crucial place in comprehending the fundamental laws of the Universe. Within the Standard Model of particle physics, accounting for the observed Parity (P) and CP violation, C being Charge-conjugation symmetry, in the weak sector is essential for building a consistent model of particle physics. Additionally, to account for the observed baryon asymmetry in the current cosmological setting, the early Universe must have violated C, CP, and Baryon number (B). In cosmology, ongoing investigations are being carried out to examine parity-violating signals, both in theory and through different kinds of observations.
Possible hints of parity violation in the Cosmic Microwave Background (CMB) emerge at a significance level of $3.6\sigma$ within the so-called ``cosmic birefringence'' effect \cite{Komatsu_Eskilt_2022,PhysRevLett.125.221301,Gruppuso_2016,PhysRevLett.128.091302}. From a theoretical point of view, this effect can be explained by means of a modification to electromagnetism through an axion-photon coupling \cite{Lue_Kamionkowski_1999,PhysRevD.41.1231,PhysRevD.101.123529,Greco_2022,Greco_2023,greco2024newsolutionobservedisotropic}.

Furthermore, in large-scale structure (LSS) analyses there is an ongoing discussion regarding possible hints of parity violation in the galaxy four-point correlation function (4PCF). Recently, two groups \cite{hou_measurement_2022,philcox2021detection} 
have reported hints of parity violation in the galaxy 4PCF within the final data release of the BOSS galaxy
survey \cite{BOSS_GALAXY_SURVEY_2015}, respectively, with statistical significance as high as $7.1\sigma$ and $2.9\sigma$. 
On the other hand, a later analysis using an alternative set of mock catalogs found no support for parity violation \cite{philcox2024sample}. Other follow-up studies, such as \cite{krolewski2024}, come to the same conclusion. Moreover, no evidence of parity violation in the 4PCF in the CMB (and CMB bispectra)  \cite{Planck:2015zfm,Shiraishi:2017yrq,Planck_2019_IX,philcox2023cmb,Philcox:2023xxk,Philcox:2023ypl}\footnote{See, e.g.~\cite{Philcox_2025_1,Philcox_2025_2,Philcox_2025_3}, for more recent studies on the 4PCF (including parity violating features) in the CMB, which extend on previous CMB trispectra analyses, e.g~\cite{Smith:2015uia,Planck:2013jfk,Planck:2015zfm,Planck_2019_IX}.} and in the public SDSS Data Release 16 Lyman-$\alpha$ forest data \cite{adari2024searchingparityviolationsdss} has been found.
As noted in \cite{Shiraishi_2016} (see also \cite{Slepian_test_cosmological_parity_violation,slepian_cahn_isotropic_2020}), the 4PCF \cite{slepian_philcox_efficient_2022} represents the lowest order statistics capable of exhibiting parity violation within the scalar sector, for example, through the use of correlators of the gauge-invariant \cite{bruni_perturbations_1997,Matarrese_1998}  curvature perturbation $\zeta(t,\mathbf{x})$ \cite{kodama_cosmological_1984}. 

Although the first claims have been questioned, there is still considerable interest in investigating parity violation in large-scale structures. Firstly, these analyses have led to significant advancements in data analysis techniques (e.g.~\cite{Jeong_2012,Masui_2017,zhu2024systematicanalysisparityviolatingmodes}). For example, precise algorithms for performing such analyses were developed in \cite{Slepian_test_cosmological_parity_violation}, based on innovative methods \cite{10.1093/mnras/stv2119,slepian_cahn_isotropic_2020,10.1093/mnras/stab3025,Analytic_Gaussian} for the 4PCF. Other examples include the introduction of novel observables for studying parity violation in LSS analyses, such as Parity-Odd Power Spectra \cite{Caravano} in which the parity-odd trispectrum is compressed into simpler statistics computationally fast to construct.\footnote{For analyses of the parity-even part of the trispectrum, see instead, e.g.~the techniques used in \cite{Gualdi_1,Gualdi_2,Gualdi_3}.}  
Furthermore, the analysis of the parity-odd 4PCF can provide bounds on cosmological models that are competitive if not stronger than those obtained from the CMB, see \cite{philcox_probing_2022} for an example.
From a theoretical perspective, studying models that imprint parity-violating signatures is often equivalent to studying new-physics signatures, and it is well justified given the significantly enhanced sensitivities expected from forthcoming galaxy surveys. Additionally, these theoretical studies could offer valuable insights into primordial non-Gaussianity \cite{Bartolo_2004}: a scalar field exhibiting parity violation in the 4PCF would also be non-Gaussian.

Einstein's theory of gravity, i.e.~General Relativity (GR), preserves parity leading to standard inflationary models \cite{BROUT197878,Sato_Katsuhiko,PhysRevD.23.347_guth,starobinsky_new_1980,LINDE1982389,Albrecht_Andreas_Steinhardt} in agreement with observational data \cite{WMAP_2013,Planck_VI_2018,Planck_2019_IX,Planck_2020_X}. To describe parity-violating features in LSS, we explore extensions of GR which play a relevant role during inflation.  It is possible for new terms to emerge in the Lagrangian density, some of which may violate parity. The simplest example is the four-dimensional Chern-Simons (CS) term coupled to a scalar field \cite{Lue_Kamionkowski_1999}. This term emerges in the context of string theory \cite{GREEN1984117,Polchinski:1998rr,Choi_2000,Alexander_2006,Lyth_2005,Alexander_2009,Kamada_2020}, loop quantum gravity \cite{ASHTEKAR_1989,Taveras_2008,Calcagni_2009,Gates_2009,Mercuri_2009} or Ho\ifmmode \check{r}\else \v{r}\fi{}ava-Lifshitz gravity with the so-called three-dimensional CS term \cite{Horava_2009,Horava_2017_2}.\footnote{For a general discussion on possible mechanisms producing the CS term see \cite{Alexander_2009}.} Moreover, the four-dimensional Chern-Simons operator arises naturally as a low energy effective field theory (EFT) through an expansion in curvature invariants \cite{weinberg_effective_2008}. From the viewpoint of an EFT approach, the Chern-Simons term is a leading-order correction to the Hilbert-Einstein term. Inspired by the extension of CS gravity theory, \cite{crisostomi_beyond_2018} introduced new chiral scalar-tensor (CST) theories of gravity. These theories incorporate new parity-breaking terms that involve additional derivatives of both the metric and a scalar field, compared to the Hilbert-Einstein and Chern-Simons actions. These theories were initially developed for cosmological applications concerning dark energy. However, in this context, we use them in an inflationary setting. As discussed in \cite{crisostomi_beyond_2018}, Ostrogradsky pathological modes arise when working with the full version of these new theories. To address this issue, two strategies can be employed. The first one is to treat these theories as low-energy EFTs which are valid up to the energy scale at which the pathological modes emerge. The second strategy involves adopting the unitary gauge and imposing additional constraints on the free parameters defined in the full action. Although this approach eliminates the Ostrogradsky modes, it requires interpreting these theories as new Lorentz-breaking theories. In this work, we adopt the second approach.

The physical properties and observational outcomes of the primordial power spectra and certain bispectra of these models have already been studied in the literature. The study of chirality of gravitational waves (GWs) in these observables has been performed in \cite{Takahashi_2009,Wang_2013,alexander_birefringent_2005,Satoh_2010,Bartolo_2017,creque-sarbinowski_parity-violating_2023,Bartolo_2017} for the CS theory\footnote{The Chern-Simons theory has also been studied in other contexts, including black hole physics (see, e.g.~\cite{Konno_2009,Cardoso,Chen_2010,Grumiller_2008}) and the propagation and generation of gravitational waves (see, e.g.~\cite{Wagle_2019,Yunes_2023,califano2024parityviolationgravitationalwaves,Alexander_2008,Alexander_2018,Yagi_2018}).} and in \cite{Qiao,bartolo_tensor_2021} for the CST theories. 
The amount of chirality produced by each model in the tensor (i.e.~GW) power spectrum is small and observations cannot distinguish between the two theories. As a result, the two-point statistics alone are insufficient to discriminate between the models.
From an observational point of view, it is well recognized that parity-violating primordial tensor power spectra can source non-zero parity-breaking CMB angular correlators $TB$ and $EB$ \cite{Lue_Kamionkowski_1999}. However, recent analysis suggests that the CMB angular power spectra is sensitive to parity violation in the early Universe exclusively in cases of maximal chirality, even under optimal conditions \cite{Gerbino_2016}. By cross-correlating data from multiple detectors, including both ground-based and space-based interferometers, it is, a priori, feasible to detect parity violation stemming from the primordial Universe \cite{Seto_2008,Smith_2017,Thorne_2018,Domcke_2020,Seto_2020,Orlando_2021}, i.e.~bypassing CMB and looking directly to the GWs produced during inflation. However, upcoming interferometers, such as LISA \cite{Bartolo_2016} or the Einstein telescope \cite{Maggiore_2020}, are expected to detect the primordial gravitational wave background solely for a specific range of inflationary models where primordial gravitational waves exhibit a blue tensor tilt. 
There are also additional proposals, albeit in very early stages, to detect chirality in power spectra using advanced \SI{21}{\centi\meter} survey measurements \cite{Masui_2017} and the 2D galaxy shear power spectrum \cite{Biagetti_2020}.
Considering the challenges in testing these models with only two-point statistics, theoretical analyses and detection prospects in CMB of parity-violating bispectra (arising in the tensor sector) have been performed \cite{Bartolo_2017,Bartolo_2019,bartolo_tensor_2021,Soda_2011,Maldacena_2011,Shiraishi_2011,Huang_2013,Zhu_2013,Bordin_2017,C_rdova_2017,Bordin_2020,philcox2024nongaussianityscalarsectorsearch,Akama_2024}.

Considering the current lack of direct detection of tensor modes and the aforementioned challenges in future detection of parity-violating signatures in cosmological power-spectra, it is natural to investigate parity violation within the scalar sector, such as LSS galaxy clustering or the temperature $T$ anisotropies and $E$-mode polarization in the CMB field (see \cite{Shiraishi_2016}). This is also a motivating factor for hunting for parity-breaking signatures in LSS with the 4PCF. As mentioned previously, within the scalar sector, the trispectrum, which is the Fourier transform of the 4PCF, represents the lowest order purely scalar statistics that can exhibit parity-violating features. For instance, the authors of \cite{creque-sarbinowski_parity-violating_2023} have investigated the parity-odd scalar trispectrum of the gauge-invariant curvature perturbation in an inflationary single-field slow-roll model modified by a dynamical Chern-Simons (dCS) interaction term.\footnote{For previous calculations on parity-violating trispectrum in Chern-Simons theory see \cite{Salvarese}.} In this scenario, parity violation arises due to the different propagation of the two chiral gravitational waves polarization's modes, namely the left- and right-handed ones. Their results indicate that within a single-field slow-roll inflationary model, the resulting trispectrum cannot account for the observed hints of parity violation reported e.g.~in  \cite{hou_measurement_2022}. With adjustments to such a framework, referred to as beyond dCS, such as a quasi single-field slow-roll model or a superluminal scalar sound speed, it is feasible to enhance the signal and generate a trispectrum capable of inducing parity violation in LSS. Furthermore, \cite{Cabass_2023_2} established a set of no-go theorems concerning the parity-violating scalar trispectrum within the decoupling limit of the effective field theory of inflation \cite{Cheung_2008}, under the assumptions of scale-invariance and a Bunch-Davies vacuum.\footnote{They also provide a series of yes-go examples relaxing some of the previous assumptions, considering, e.g., breaking of scale-invariance or non-Bunch-Davies initial conditions as in Ghost inflation \cite{Nima_Arkani_Hamed_2004}. 
The no-go theorems derived in \cite{Cabass_2023_2} do not apply here since the tensor mode functions we are working with are not the de Sitter ones.} Further examples of investigation of parity-violating 4PCF during inflation is presented in \cite{Liu_2020,Niu_2023,Reinhard_2024}. \cite{inomata2024paritybreakinggalaxy4pointfunction} shows that a chiral gravitational wave background could imprint parity-violating signatures in the galaxy 4PCF. Moreover, for earlier results on parity-even inflationary scalar trispectrum, see, for example, \cite{Huang_2006,Seery_2007,Seery_2009}.

Driven by the growing interest in parity-violating physics in cosmology and recently in 4PCFs of cosmological perturbations, in this paper we present a computation of the graviton-mediated scalar trispectrum within chiral scalar-tensor theories. A single parity-violating scalar-scalar-tensor operator is used in the calculation. We employ this operator in both vertices required to form the graviton-mediated trispectrum. In this study, we focus on providing an estimate of the overall signal, rather than exploring the full shape of the graviton-mediated trispectrum. Our computation shows that a parity-breaking trispectrum can arise in the single-field slow-roll inflationary model of these theories. This work is carried out with a specific choice for the free functions of the theories, with a more general and detailed analysis to be presented in a forthcoming paper \cite{workinprogress}. The framework we adopt here is the only one capable of generating a detectable parity-violating trispectrum without modifying the single-field slow-roll model of CST theories. In alternative scenarios within this theoretical framework, it remains feasible to generate a detectable signal; however, this requires adjustments like those proposed, e.g.~in \cite{creque-sarbinowski_parity-violating_2023}, including quasi-single-field models or superluminal scalar sound speed.

The paper is organized as follows. In section \ref{Sec:Chiral_scalar-tensor_theories_of_gravity}, we briefly review the CST theories, describe the formalism used in our analysis, and derive the power spectrum statistics associated with these theories. In section~\ref{Sec:Parity-violating-trispectrum}, we present the computation of the graviton-mediated trispectrum and estimate the signal-to-noise ratio of the parity-violating part of this trispectrum, while in section~\ref{Sec:Conclusions} we draw our conclusions.

\section{Chiral scalar-tensor theories of
gravity}
\label{Sec:Chiral_scalar-tensor_theories_of_gravity}

In this section, following the idea of an EFT approach to modify the standard slow-roll single-field model of inflation  we introduce the so-called chiral scalar-tensor theories of gravity proposed in \cite{crisostomi_beyond_2018}. 
Within these theories, we introduce covariant parity-breaking terms which have more derivatives with respect to both the Hilbert-Einstein term and the Chern-Simons term. These theories lead to equations of motion with higher-order derivatives, and the Ostrogradsky's thereom states that this may make the theory pathological with the appearance of ghost's modes \cite{pons_ostrogradskis_1989,Woodard_1,Woodard2007}. However, under suitable conditions, it is possible to
show that this occurrence is not verified in CST theories of modified gravity \cite{crisostomi_beyond_2018}. 
The action of the theories we are considering has the following form
\begin{align}
    S_{\mathrm{PV}} &=  \int \mathrm{d}^4 x\,\sqrt{-g}\left[\frac{M_{\mathrm{Pl}}^2}{2}R-\frac{1}{2}g^{\mu\nu}\nabla_{\mu}\phi\nabla_{\nu}\phi - V(\phi)+\mathcal{L}_{\mathrm{PV}}\right]\,  , 
    \label{Section_1.1*):Global_action}
\end{align}
where, $\nabla_{\mu}$ is the covariant derivative, $g = \det[g_{\mu\nu}]$, $M_{\mathrm{Pl}} = \left(8\pi G\right)^{-\frac{1}{2}}$ is the reduced Planck mass, $R$ is the Ricci scalar, $\phi$ is the inflaton field which is not minimally coupled to gravity and $\mathcal{L}_{\mathrm{PV}}$ is the Lagrangian which contains parity-violating operators  defined as the sum of two pieces:
\begin{align}
    \mathcal{L}_{\mathrm{PV}} = \mathcal{L}_{\mathrm{PV}1} + \mathcal{L}_{\mathrm{PV}2}\,.
    \label{Eq:Chiral_scalar_tensor_theories)LPV}
\end{align}
These are the parity-violating Lagrangians which  include parity-breaking operators involving first and second derivatives of the inflaton field $\phi$, respectively.
The first reads \cite{crisostomi_beyond_2018}
\begin{align} 
    \mathcal{L}_{\rm PV1} =& \sum_{A=1}^4  a_{A} L_{A}\, ,
    \label{Eq:Chiral_scalar_tensor_theories)LPV1}
\end{align}
where 
\begin{align}
    L_1 &= \frac{\tilde{\varepsilon}^{\mu\nu\alpha \beta}}{M_{\mathrm{Pl}}^4} R_{\alpha \beta \rho \sigma} R_{\mu \nu\; \lambda}^{\; \; \;\rho} \phi^\sigma \phi^\lambda  \, , \qquad \qquad\qquad \qquad L_3 = \frac{\tilde{\varepsilon}^{\mu\nu\alpha \beta}}{M_{\mathrm{Pl}}^4} R_{\alpha \beta \rho \sigma} R^{\sigma}_{\;\; \nu} \phi^\rho \phi_\mu \, ,\nonumber\\
    L_2 &=  \frac{\tilde{\varepsilon}^{\mu\nu\alpha \beta}}{M_{\mathrm{Pl}}^4} R_{\alpha \beta \rho \sigma} R_{\mu \lambda }^{\; \; \;\rho \sigma} \phi_\nu \phi^\lambda \, , \qquad \qquad\qquad \qquad  L_4 =  \frac{\tilde{\varepsilon}^{\mu\nu\rho\sigma}}{M_{\mathrm{Pl}}^4} R_{\rho\sigma \alpha\beta} R^{\alpha \beta}_{\;\;\;\; \mu\nu} \phi^\lambda \phi_\lambda \, ,
\label{opLPV1}
\end{align}
where $R^{\alpha}{}_{\beta \rho \sigma}$ is the Riemann tensor, $\phi^{\,\mu} = \nabla^\mu \phi$ and, $\tilde{\varepsilon}_{\rho \sigma \alpha \beta}$ is the Levi-Civita tensor density defined as 
 \cite{carroll_2019}, 

\begin{align}
    \tilde{\varepsilon}_{\rho \sigma \alpha \beta}\equiv\frac{{\varepsilon}_{\rho \sigma \alpha \beta}}{\sqrt{-g}}\,,
\end{align}
where ${\varepsilon}_{\rho \sigma \alpha \beta}$ is the Levi-Civita symbol.
The couplings $a_A$ in \eqref{Eq:Chiral_scalar_tensor_theories)LPV1} are dimension-less free functions of the scalar field and its kinetic term, i.e.~$a_A = a_{A}(\phi, \phi^\mu \phi_\mu)$.
In order to avoid the Ostrogradsky modes, we have to work in the unitary gauge, i.e.~$\delta\phi(x^{\mu})=0,$ setting \cite{crisostomi_beyond_2018}
\begin{equation} 
    4a_1+2 a_2+a_3 +8 a_4=0 \, ,
    \label{Eq:Chiral_scalar_tensor_theories)constraint_a}
\end{equation}
from which we understand that we have only three independent functions. 
The second Lagrangian $\mathcal{L}_{\rm PV2}$ reads \cite{crisostomi_beyond_2018}
\begin{align}
    \mathcal{L}_{\rm PV2} &= \sum_{A=1}^7 b_{A} M_{A}\, ,
     \label{Eq:chiral_scalar_tensor_theories)LPV2}
\end{align}
where
\begin{align}
    M_1 &= \frac{\tilde{\varepsilon}^{\mu\nu \alpha \beta}}{M_{\mathrm{Pl}}^5} R_{\alpha \beta \rho\sigma} \phi^\rho \phi_\mu \phi^\sigma_\nu \, , \qquad \qquad\qquad \qquad \hspace{0.35cm} M_4 = \frac{\tilde{\varepsilon}^{\mu\nu \alpha \beta}}{M_{\mathrm{Pl}}^8} R_{\alpha \beta \rho\sigma} \phi_\nu \phi^\rho{}_\mu \phi^\sigma{}_\lambda \phi^\lambda, \nonumber\\
    M_2 &= \frac{\tilde{\varepsilon}^{\mu\nu \alpha \beta}}{M_{\mathrm{Pl}}^4} R_{\alpha \beta \rho\sigma} \phi^\rho{}_\mu \phi^\sigma{}_\nu \, , \qquad \qquad\qquad \qquad\,\,\,\,\,\, M_5 = \frac{\tilde{\varepsilon}^{\mu\nu \alpha \beta}}{M_{\mathrm{Pl}}^8} R_{\alpha \rho\sigma \lambda } \phi^\rho \phi_\beta \phi^\sigma{}_\mu \phi^\lambda{}_\nu,  \nonumber\\
    M_3 &= \frac{\tilde{\varepsilon}^{\mu\nu \alpha \beta}}{M_{\mathrm{Pl}}^8} R_{\alpha \beta \rho\sigma} \phi^\sigma \phi^\rho{}_\mu \phi^\lambda{}_\nu \phi_\lambda\, , \qquad \qquad\qquad\,\,\,\,\,   M_6 = \frac{\tilde{\varepsilon}^{\mu\nu \alpha \beta} R_{\beta \gamma}}{M_{\mathrm{Pl}}^8} \phi_\alpha \phi^\gamma{}_\mu \phi^\lambda{}_\nu \phi^\lambda, \nonumber\\
    M_7 &= \frac{(\Box \phi)}{M_{\mathrm{Pl}}^3} M_1 \, ,
\label{opLPV2}
\end{align}
where  $\phi^{\sigma}{}_\nu = \nabla^\sigma \nabla_\nu \phi$ and $b_A$ are dimension-less free functions of the scalar field and its kinetic term, i.e.~$b_A = b_{A}(\phi, \phi^{\,\mu} \phi_\mu)$. To avoid the Ostrogradsky pathological modes, we work in the unitary gauge with the following conditions \cite{crisostomi_beyond_2018}
\begin{equation} 
     b_7 = 0 \, , \qquad  b_6 = 2(b_4 + b_5) \,,  \qquad b_2 =-\frac{A_\ast^2(b_3 -b_4)}{2} \, ,
     \label{Eq:Chiral_scalar_tensor_theories)constraint_b}
\end{equation}
where $A_{\ast}\equiv\dot{\phi}(t)/N$ with $\dot{}\equiv\partial/\partial t$ and $N$ is the lapse function of the spacetime \cite{wald_general_1984,ADM_formalism}. We consider the Lagrangian in eq.~\eqref{Eq:Chiral_scalar_tensor_theories)LPV}, subject to the constraints given in eqs.~\eqref{Eq:Chiral_scalar_tensor_theories)constraint_a} and \eqref{Eq:Chiral_scalar_tensor_theories)constraint_b}, as the fundamental Lagrangian with which we are working. According to the discussion in \cite{crisostomi_beyond_2018,bartolo_tensor_2021} and the content presented in section \ref{Sec:introduction}, this approach is equivalent to treating these theories as new Lorentz-breaking theories.

\subsection{Adopted formalism}
\label{Sec:chiral_scalar_tensor)The actions_in_the_ADM_framework}
To perform the original computations regarding the graviton-mediated trispectrum of the gauge-invariant curvature perturbation $\zeta$, we work in the Arnowitt-Deser-Misner (ADM) formalism of the metric \cite{ADM_formalism} adopting the comoving gauge and perturbing a spatially flat FLRW metric.
In Cartesian coordinates, the line element reads
\begin{align}
    \mathrm{d}s^{2}  =
       -(N^{2}-N_{i}N^{i}) \mathrm{d}t^{2} + N_{i} \mathrm{d}x^{i} \mathrm{d}t +
         g_{ij}  \mathrm{d}x^{i} \mathrm{d}x^{j}\,,
    \label{Eq:Chiral_scalar_tensor)line_element}
\end{align}
where $N$ is the lapse function and, $N_{i}$ is the shift vector which, based on the Helmholtz decomposition \cite{stewart_perturbations_1990}, can be expressed as the sum of a curl-free vector and a divergence-free vector as
\begin{align}
    N_i \equiv \partial_{i}\psi + E_{i}\,, \quad \partial_{i}E^{i} = 0\,,
\end{align}
where we adopt the following convention for three-dimensional spatial partial derivatives
$\partial_i\psi = \partial\psi/\partial x^i$. Moreover, $g_{ij}$ is the spatial metric which is decomposed as
\begin{align}
    g_{ij} &= a(t)^{2}e^{2\zeta} \exp{\gamma}_{ij}\,,\quad  \exp{\gamma}_{ij}= \left(\delta_{ij} + \gamma_{ij} + \frac{1}{2}\gamma_{ik}\gamma^{k}{}_{j} +  {\dots}\right)\,,\quad
    \gamma^{i}{}_{i} = 0\,, \quad \partial_{i}\gamma^{i}{}_{j}=0\,,
\end{align}
where $a(t)$ is the scale factor of the Universe,  $\zeta$ is the gauge-invariant curvature perturbation  and, $\gamma_{ij}$ represents the transverse and traceless tensor perturbations \cite{Salopek:1990jq,maldacena_non-gaussian_2003}. 
According to the Hamiltonian analysis performed in \cite{crisostomi_beyond_2018}, we can write the parity-violating Lagrangian densities in the ADM formalism as
\begin{align}\label{Eq:Chiral_scalar_tensor)PV1-action-ADM}
    \nonumber
    \sqrt{-g}\,\mathcal{L}_{\mathrm{PV}1}
    &=\frac{4\epsilon }{N}\left(\frac{H}{M_\mathrm{Pl}}\right)^{2}\varepsilon^{ijl}\biggl[ 2(2a_1+a_2+4a_4)\biggl(KK_{mi}D_lK^{\thinspace m}{}_j\,+\,\prescript{(3)}{\noindent}{R_{mi}}D_lK^{\thinspace m}{}_{j}\,+\\&-\,K_{mi}K^{mn}D_lK_{jn}\biggr)
    \,-\,(a_2+4a_4)\left(2K_{mi}K^{\thinspace n}{}_jD_nK^{\thinspace m}{}_l\,+\,\prescript{(3)}{\noindent}{R_{jlm}}^nD_nK^{\thinspace m}{}_i\right)\biggr] \, ,\\
    \label{Eq:Chiral_scalar_tensor)PV2-action-ADM}
    \nonumber
    \sqrt{-g}\mathcal{L}_{\mathrm{PV}2} &= \frac{1}{N^{4}}\left(\frac{H}{M_{\mathrm{Pl}}}\right)^{2}\varepsilon^{ijl}\biggl\{ 2N\biggl[(2\epsilon)^{\frac{3}{2}}b_{1}N H K_{mi}D_{l}K^{m}{}_{j} \nonumber\,+\,\\&+\, 4\epsilon^2\left(\frac{H}{M_{\mathrm{Pl}}}\right)^2\left(b_{4} + b_{5}-b_{3}\right)K_{mi}K^{n}{}_{j}D_{n}K^{m}{}_{l}\biggr]\,+\,\nonumber\\&+\,4\epsilon^2\left(\frac{H}{M_{\mathrm{Pl}}}\right)^2\biggl(b_{3}{}^{(3)}R_{jlm}{}^{n}K^{m}{}_{i}D_{n}N\,-\,2(b_{4}+b_{5}){}^{(3)}R_{ml}K^{m}{}_{j}D_{i}N\biggr)\biggr\}\,,
\end{align}
where we have used the slow-roll condition $\dot{\phi}^2 \approx 2 \epsilon M_{\mathrm{Pl}}^2H^2$ with $\epsilon$ being the first slow-roll parameter \cite{Lidsey:1995np} and $H$ the inflationary Hubble parameter, respectively. Furthermore, we have introduced the three-dimensional covariant derivative operator $D_{l}$ \cite{wald_general_1984}, as well as the three-dimensional Riemann and Ricci tensors, $\prescript{(3)}{\noindent}{R^{k}{}_{jli}}$ and $\prescript{(3)}{\noindent}{R_{ij}}$, and the extrinsic curvature tensor 
\begin{align}
    K_{ij} = \frac{1}{2N}\left(\dot{g}_{ij} - D_{j}N_{i} - D_{i}N_{j}\right)\,.
\end{align}

The lapse and shift functions are auxiliary fields in standard gravity, and they can be removed by solving the Equations of Motion (EoMs). However, since we have modified the gravity sector, it is necessary to determine whether these fields become dynamical. Since we will need to expand the action up to third order\footnote{To compute the graviton-mediated trispectrum we need the trilinear scalar-scalar-tensor vertices.} we are only interested in first-order solutions to the constraint equations for the lapse and shift functions \cite{Bartolo_2017,bartolo_tensor_2021,maldacena_non-gaussian_2003,Pajer2017}. Thus, we need to discuss all the bilinear terms we can construct at the Lagrangian level to determine whether modifications arise at linear level in the EoMs. The only non-vanishing bilinears we can construct are of the tensor-tensor and vector-vector type which, up to space-independent multiplicative factors, read
\begin{align}
    \epsilon^{ijk}\gamma_{li}{}^{,m}\gamma^{l}{}_{k,mj}\,, \hspace{40pt}
    \epsilon^{ijk}N_{i,j}N_{k}
    \label{Eq:CS)non_zero_terms_bilinear_lagrangian}\,,
\end{align}
where we use the following convention for spatial derivatives $\partial_jN_i = N_{i,j}$.
This shows us that the equation for the lapse function, $N$, is unchanged with respect to standard GR and remains a non-dynamical Degree of Freedom (DOF). Furthermore, given $\dot{E}_{i}$ is not present in either eq.~\eqref{Eq:Chiral_scalar_tensor)PV1-action-ADM} or eq.~\eqref{Eq:Chiral_scalar_tensor)PV2-action-ADM}, the shift function remains non-dynamical although its equation of motion differs from the standard case. However, these additional contributions are linear terms in the transverse vector field $E_{i}$, so, as in the case of standard gravity, we can find a solution by setting $E_{i}$ to zero \cite{maldacena_non-gaussian_2003}:
\begin{align}
    N = 1 + \frac{\dot{\zeta}}{H}\,, \qquad E^{i} = 0\,, \qquad \psi = -\frac{\zeta}{a^2H} + \alpha\,, \qquad \nabla^{2}\alpha
    \approx \epsilon \dot{\zeta}\,,
\end{align}
where we have used $\alpha$ instead of $\chi$, contrary to \cite{maldacena_non-gaussian_2003}, in order to avoid confusion with the notation for the chirality parameter we will introduce in the next section.

\subsection{The EoMs for primordial scalar and tensor modes}\label{Sec:chiral_scalar_tensor)The EoM and the Power spectrum}
In this section, we review the computations of the EoM for scalar and tensor modes.
We work in Fourier space
\begin{align}
    \zeta(\mathbf{x},\tau) &=\int\frac{\mathrm{d}^3k}{(2\pi)^3}u_\zeta(\mathbf{k},\tau) e^{i\mathbf{k}\cdot\mathbf{x}}\,,\\
    \gamma_{ij}(\mathbf{x},\tau) &=\int\frac{\mathrm{d}^3k}{(2\pi)^3}\sum_{s=L,R}\gamma_s(\mathbf{k},\tau)\epsilon^{(s)}_{ij}(\mathbf{k}) e^{i\mathbf{k}\cdot\mathbf{x}}\, ,
\end{align}
where $\gamma_s(\mathbf{k},\tau)$ and $u_\zeta(\mathbf{k},\tau)$ are the Fourier components of the primordial tensor and scalar modes, respectively, $s = R/L$ being the polarization index, while $\epsilon^{(s)}_{ij}(\mathbf{k})$ is the polarization tensor in the chiral basis \cite{Bartolo_2017} of \textit{left} and \textit{right} circular polarization states (see appendix~\ref{Polarization tensors}). 
Applying the same reasoning as in the previous section we can conclude that in eq.~\eqref{Eq:Chiral_scalar_tensor)PV1-action-ADM} and eq.~\eqref{Eq:Chiral_scalar_tensor)PV2-action-ADM} we do not have any scalar bilinears contribution. Thus,
 the scalar equation of motion remains unchanged. Imposing the Bunch-Davies initial condition and considering the lowest order in slow-roll parameters, we obtain the scalar mode function (e.g.~\cite{Bartolo_2004}),
\begin{align}
    u_{\zeta}(k,\tau) 
    = \frac{iH}{\sqrt{4\epsilon k^{3}}M_{\mathrm{Pl}}}\left(1+ik\tau\right)e^{-ik\tau}\,, 
\end{align}
which leads to
\begin{align}   
    P_{\zeta} = \frac{1}{4\epsilon k^3}\left(\frac{H}{M_{\mathrm{Pl}}}\right)^2\,,
    \label{Eq:scalar_power_spectrum}
\end{align}
where $P_{\zeta}$ is the primordial scalar power spectrum on super-horizon scales \cite{Bartolo_2004}. We work at lowest order in slow-roll parameters, because the parity-violating graviton-mediated trispectrum we are going to compute in section~\ref{Sec:Parity-violating-trispectrum} still yields a non-zero contribution.

Concerning tensor modes, as shown in \cite{bartolo_tensor_2021}, each parity-violating Lagrangian, i.e., $\mathcal{L}_{\mathrm{PV}1}$ and $\mathcal{L}_{\mathrm{PV}2}$, could modify the EoM depending on the choice of free functions $a_A$ and $b_A$. Combining the two theories, the second-order action in tensor modes reads \cite{bartolo_tensor_2021}
\begin{equation}
    \label{PV2-quadratic}
    S^{(2)}_{\gamma\gamma}=\sum_{s=L,R}\int\mathrm{d}\tau\int\frac{\mathrm{d}^3k}{(2\pi)^3}\left[{A}_{T,s}^2|\gamma'_s(\mathbf{k},\tau)|^2-B_{T,s}^2k^2|\gamma_s(\mathbf{k},\tau)|^2\right] \, ,
\end{equation}
where we have switched to conformal time, $\mathrm{d}t = a(\tau)\mathrm{d}\tau $, we have used $'\equiv \partial/\partial\tau$, and we have introduced
\begin{align}
    \label{ATMPv2}
    {A}_{T,s}^2\equiv\frac{M_{\mathrm{Pl}}^2}{2}a^2
    \left(1-\lambda_s\frac{k_{\mathrm{phys}}}{M_{\mathrm{PV}2}}-\lambda_s\frac{k_{\mathrm{phys}}}{M_{\mathrm{PV}1}}\right), \quad
    B_{T,s}^2\equiv\frac{M_{\mathrm{Pl}}^2}{2}a^2\left[1-\frac{4}{M_{\mathrm{Pl}}^6}\frac{\dot{\phi}^2}{a}(\dot{f}+\dot{g})\lambda_sk\right]\,,
\end{align}
where we have introduced two mass scales defined as
\begin{align}
    M_{\mathrm{PV1}}\equiv\frac{M_{\mathrm{Pl}}^6}{8}\frac{1}{\dot{\phi}^2}\frac{1}{(f+g)H}, \quad M_{\mathrm{PV2}}\equiv\frac{M_{\mathrm{Pl}}}{2}\left(\tilde{b}_1-b\frac{H}{M_{\mathrm{Pl}}}\right)^{-1}\,, 
\end{align}
with 
\begin{equation}
    \left(f+g\right) \equiv \left(a_1+a_2+4a_4\right)\, ,\quad
    \tilde{b}_1\equiv\frac{\dot{\phi}^3}{M_{\mathrm{Pl}}^6}b_1 \, ,\quad b\equiv \frac{\dot{\phi}^4}{M_{\mathrm{Pl}}^8}(b_4+b_5-b_3) \, .
\end{equation}
The expressions for ${A}_{T,s}^2$ and $B_{T,s}^2$ are different from those presented in \cite{bartolo_tensor_2021}. In that work, the analysis focuses on models where the action consists of the standard GR action plus either $\mathcal{L}_{\mathrm{PV}1}$ or $\mathcal{L}_{\mathrm{PV}2}$, with the aforementioned quantities referring to modifications introduced by one of these two parity-violating theories. However, in our case, as we will see (section~\ref{Sec:Parity-violating-trispectrum}), both Lagrangians need to be considered. As a result, we take into account the modifications introduced at quadratic order in tensor perturbations simultaneously, resulting in eq.~\eqref{PV2-quadratic}.

As discussed in \cite{bartolo_tensor_2021} the right-handed graviton modes, i.e.~$\lambda_R=+1$, can acquire a negative kinetic term if $k_{\mathrm{phys}}>M_{\mathrm{PV1}}$ or $k_{\mathrm{phys}}>M_{\mathrm{PV2}}$. 
In such cases, these modes become unstable, potentially leading to critical issues within the theory, such as a breaking of unitarity or the propagation of negative energy modes forward in time. To avoid such circumstances, we introduce a cut-off scale for the physical momentum, $k_{\mathrm{phys}}$, denoted as $\Lambda \leq M_{\mathrm{PV1/2}}$. At the beginning of inflation, we restrict to gravitons for which  $k_{\mathrm{phys}}<\Lambda$ and $k_{\mathrm{phys}} \gg H$ \cite{bartolo_tensor_2021}. This is equivalent to work within the effective field theory limit of the theory enforcing
\begin{align}
    \chi_1 \equiv \frac{H}{M_{\mathrm{PV}1}} \ll 1\, , \quad \chi_2 \equiv \frac{H}{M_{\mathrm{PV}2}} \ll 1\,,
    \label{Eq:chiral_condition}
\end{align}
where we have introduced two \textit{chirality} parameters, $\chi_1$ and $\chi_2$, in analogy to what is done, e.g., in \cite{bartolo_tensor_2021,Bartolo_2017,creque-sarbinowski_parity-violating_2023}.
The chiral conditions in eq.~\eqref{Eq:chiral_condition} can be expressed as upper bounds on the values of the free functions, $a_A$ and $b_A$. For example, the condition for $\chi_1$ can be written as
\begin{align}
    (f+g) \ll 2.2 \times10^{18}\left(\frac{\SI{1e14}{\giga\electronvolt}}{H}\right)^4\left(\frac{10^{-2}}{\epsilon}\right)\,,
        \label{Eq:Chiral_limit_a_s}
\end{align}
where $\epsilon$ is the slow-roll parameter. 
Working with conditions like the one in eq.~\eqref{Eq:Chiral_limit_a_s} might appear to limit the potential use of free functions to enhance the signal of the graviton-mediated trispectrum. However, we impose the following conditions
\begin{align}
    a_1 + a_2 + 4a_4 = 0\, , \quad \quad b_1 -\sqrt{2\epsilon} \left(\frac{H}{M_{\mathrm{Pl}}}\right)^2(b_4+b_5-b_3)\neq0\,,
\label{Eq:specific_parameters_configuration}
\end{align}
which basically correspond to set $M_{\mathrm{PV}1}\equiv\infty$, i.e.~$\chi_1 = 0$, while keeping $M_{\mathrm{PV}2}$ finite, i.e.~\mbox{$\chi_2\neq0$}.\footnote{Please note that we can decide to work with $M_{\mathrm{PV}1}$ finite and $M_{\mathrm{PV}2}=\infty$. The idea is to eliminate the constraints imposed by chiral conditions on the free functions of one Lagrangian density while retaining the chirality introduced in the tensor EoM by the other.} This allows us to get rid off this bound for the parameters $a_A$ while preserving the chirality in graviton propagation because of the modifications introduced by $\mathcal{L}_{\mathrm{PV}2}$ in the tensor EoM. In this setup, $\mathcal{L}_{\mathrm{PV}1}$ does not modify the equations of motion of scalar and tensor perturbations, while only adding new interaction terms whose amplitude is not bounded by any chiral conditions such as that in eq.~\eqref{Eq:chiral_condition}.
The only constraint is related to maintaining the validity of the perturbative treatment (see appendix~\ref{Sec:perturbativity_ratio}
). As we will see in section~\ref{Sec:Parity-violating-trispectrum}, this provides a recipe to have a detectable trispectrum in single-field slow-roll inflation within chiral scalar-tensor theories of modified gravity.\footnote{All the other possibilities will be presented in \cite{workinprogress}.}

We are not going to solve the EoM\footnote{See eqs.~(3.24), (3.25), (3.26), (3.27), and (3.28) in \cite{bartolo_tensor_2021} for an explicit expression.} since this is unnecessary for what we will use in the next section. However, this can be done \cite{Qiao,bartolo_tensor_2021} and the resulting power spectra for tensor perturbations at leading order in slow-roll are as follows
\begin{align}
    P_T^L= \frac{P_T}{2} \exp\left[\frac{\pi}{16}\frac{H}{M_{\mathrm{PV}2}}\right] \, ,\qquad P_T^R= \frac{P_T}{2} \exp\left[-\frac{\pi}{16}\frac{H}{M_{\mathrm{PV}2}}\right] \, ,
    \label{Eq:tensor_PS}
\end{align}
where $T$ stands for tensor, $R$ and $L$ denote right and left, respectively, and $P_T$ refers to the tensor power spectrum of GR \cite{Bartolo_2004}
\begin{align}
    P_{T} = \frac{4}{k^3}\left(\frac{H}{M_{\mathrm{Pl}}}\right)^2.
\end{align}

\section{Parity-violating graviton-mediated trispectrum}\label{Sec:Parity-violating-trispectrum}
In this section, we present an original computation of the connected graviton-mediated trispectrum of the gauge-invariant curvature perturbation $\zeta$, i.e.
\begin{align}
    \vcenter{\hbox{\begin{tikzpicture}[ scale=0.5]
      \begin{feynman}
    \vertex [label=left:\(\zeta(\mathbf{K}_{I}^{1})\), square dot] (f1) at (0,1.5) {};
    \vertex [dot, minimum size=3pt] (f2) at (1.5,0) {};
    \vertex [dot, minimum size=3pt] (f3) at (3.5,0) {};
    \vertex [square dot, label=right:\(\zeta(\mathbf{K}_{I}^{3})\)] (f4) at (5,1.5) {};
    \vertex [square dot, label=left:\(\zeta(\mathbf{K}_{I}^{2})\)] (p1) at (0,-1.5) {};
    \vertex [square dot, label=right:\(\zeta(\mathbf{K}_{I}^{4})\)] (p2) at (5,-1.5) {};
    \diagram* {
      (f1) -- [plain] (f2) -- [boson, edge label=\(\gamma_s(\mathbf{k}_{I})\)] (f3) -- [plain] (f4),
      (f2) -- [plain] (p1),
      (f3) -- [plain] (p2),
    };
  \end{feynman}
\end{tikzpicture}}}
\label{Eq:Parity_violating_trispectrum)The_graviton-mediated_trispectrum}
\end{align}
This represents an example of parity-violating trispectrum within the framework of the single-field slow-roll model of inflation in chiral scalar-tensor theories of modified gravity. A more thorough and detailed examination of all possibilities to generate a parity-violating trispectrum within these theories and their modifications will be presented in \cite{workinprogress}. 
Before delving into the computations, we want to highlight a few points:
\begin{itemize}
    \item  For our purposes it is sufficient to work at the zero order in the slow-roll parameters.
    \item We are computing the scalar trispectrum because it is the lowest scalar correlator capable of exhibiting parity-violating signatures \cite{slepian_cahn_isotropic_2020}.
    \item We focus on the graviton-mediated trispectrum, as it represents a theoretically interesting case in which parity violation arises in the scalar sector due to the effects of parity-violating tensor modes. 
    Moreover, let us notice that it constitutes the only parity-violating contribution to the scalar trispectrum when working with constant couplings $a_A$.\footnote{Let us point out that in section~\ref{Sec:the_computation}, we use a time-dependent coupling (see eq.~\eqref{Eq:choice_a_1}) to make an estimate of the calculation using trispectrum shapes that have already been computed in existing literature \cite{creque-sarbinowski_parity-violating_2023,Seery_2009}. Dimensional analysis suggests that this choice would not modify the overall estimate of the signal-to-noise ratio of the parity-violating part of the trispectrum we compute in section \ref{Sec:the_computation} with respect to the case in which constant coupling functions are considered. It would have been possible to perform the calculation using constant $a_A$ functions. However, to simplify the computation, we assume that specific time dependence.
    }
    \item  In Fourier space, only the \textit{imaginary} part of the trispectrum could carry parity-violating signatures \cite{Shiraishi_2016,slepian_cahn_isotropic_2020}.
\end{itemize}

\subsection{General framework for operating within the \textit{In-In} formalism}\label{Section:General framework for operating within the In-In formalism}
The calculations are carried out by using the Schwinger-Keldysh (SK) diagrammatic rules \cite{chen_schwinger-keldysh_2017} within the framework of the \textit{In-In} formalism \cite{weinberg_quantum_2005}.
We outline the key steps necessary to perform the calculations within this framework and clarify the adopted notation. The diagrammatic rules of SK outlined in \cite{chen_schwinger-keldysh_2017}, when applied to our setup, are as follows:
\\
\noindent 
\textbf{First step:}
The initial step involves identifying all the diagrams contributing, at the lowest order in slow-roll, to the graviton-mediated scalar trispectrum. Therefore, we need to expand the full Lagrangian eq.~\eqref{Section_1.1*):Global_action} to write down all the possible scalar-scalar-tensor vertices. Subsequently, we consider all the diagrams that we can construct with the computed vertices. \\
\noindent\textbf{Second step:}
For each vertices combination, we assign one of the four tri-momenta $\mathbf{k}_1,$ $\mathbf{k}_2,$ $\mathbf{k}_3,$ $\mathbf{k}_4$ to each external leg, and then examine the complete set of distinct combinations of labelings. As in standard quantum field theory (QFT), we have three channels, i.e.~$s$, $u$ and $t$ channel. Since the computations for the three channels are identical, we introduce a more convenient notation in which we label the external momenta for a generic $I=s/t/u$ channel in the following way
\begin{align}
    \mathbf{K}_{I} = (\mathbf{K}_{I}^{1}, \mathbf{K}_{I}^{2},\mathbf{K}_{I}^{3},\mathbf{K}_{I}^{4})\,,
\end{align}
where $\mathbf{K}_{I}$ is a ordered list of the momenta involved in the specific channel and $\mathbf{K}_{I}^{l}$ with $l=1/2/3/4$ refers to the ``$l$'' tri-momentum in $I$ channel.
Thus, we can write 
\begin{align}
    \mathbf{K_{s}}
    \,=\, \left(\mathbf{k}_1, \mathbf{k}_2, \mathbf{k}_3, \mathbf{k}_4\right)\,,\quad
    \mathbf{K_{u}}\, =\, \left(\mathbf{k}_1, \mathbf{k}_3, \mathbf{k}_2, \mathbf{k}_4\right)\,,\quad
    \mathbf{K_{t}}\,= \,\left(\mathbf{k}_1, \mathbf{k}_4, \mathbf{k}_2, \mathbf{k}_3\right)\,.
\end{align}
Due to momentum conservation at each vertex, we can write the global momentum conservation as 
\begin{align}
    \mathbf{K}_{I}^{1} + \mathbf{K}_{I}^{2} = \mathbf{k}_{I} =\mathbf{K}_{I}^{3} + \mathbf{K}_{I}^{4}\,,
\end{align}
where $\mathbf{k}_{I}$ is the tri-momentum of the tensor propagator. 

\noindent\textbf{Third step:}
In the SK formalism, each vertex comes as either $+$, black dot, or $-$, white dot, vertex. This is related to the fact that both the time-ordered and the anti-time-ordered operators are at work in \textit{In-In} formalism \cite{chen_schwinger-keldysh_2017}. We have four possible combinations for each diagram:
\begin{align}
    \vcenter{\hbox{\begin{tikzpicture}[ scale=0.5]
      \begin{feynman}
    \vertex [square dot] (f1) at (0,1.5) {};
    \vertex [dot, minimum size=3pt] (f2) at (1.5,0) {};
    \vertex [dot, minimum size=3pt] (f3) at (3.5,0) {};
    \vertex [square dot] (f4) at (5,1.5) {};
    \vertex [square dot] (p1) at (0,-1.5) {};
    \vertex [square dot] (p2) at (5,-1.5) {};
    \diagram* {
      (f1) -- [plain] (f2) -- [boson] (f3) -- [plain] (f4),
      (f2) -- [plain] (p1),
      (f3) -- [plain] (p2),
    };
  \end{feynman}
\end{tikzpicture}}}
\quad
\vcenter{\hbox{\begin{tikzpicture}[ scale=0.5]
      \begin{feynman}
    \vertex [square dot] (f1) at (0,1.5) {};
    \vertex [dot, minimum size=3pt] (f2) at (1.5,0) {};
    \vertex [dot, fill=white, minimum size=3pt] (f3) at (3.5,0) {};
    \vertex [square dot] (f4) at (5,1.5) {};
    \vertex [square dot] (p1) at (0,-1.5) {};
    \vertex [square dot] (p2) at (5,-1.5) {};
    \diagram* {
      (f1) -- [plain] (f2) -- [boson] (f3) -- [plain] (f4),
      (f2) -- [plain] (p1),
      (f3) -- [plain] (p2),
    };
  \end{feynman}
\end{tikzpicture}}}
\quad
\vcenter{\hbox{\begin{tikzpicture}[ scale=0.5]
      \begin{feynman}
    \vertex [square dot] (f1) at (0,1.5) {};
    \vertex [dot, minimum size=3pt] (f2) at (1.5,0) {};
    \vertex [dot, fill=white, minimum size=3pt] (f3) at (3.5,0) {};
    \vertex [square dot] (f4) at (5,1.5) {};
    \vertex [square dot] (p1) at (0,-1.5) {};
    \vertex [square dot] (p2) at (5,-1.5) {};
    \diagram* {
      (f1) -- [plain] (f2) -- [boson] (f3) -- [plain] (f4),
      (f2) -- [plain] (p1),
      (f3) -- [plain] (p2),
    };
  \end{feynman}
\end{tikzpicture}}}
\quad
\vcenter{\hbox{\begin{tikzpicture}[ scale=0.5]
      \begin{feynman}
    \vertex [square dot] (f1) at (0,1.5) {};
    \vertex [dot, fill=white, minimum size=3pt] (f2) at (1.5,0) {};
    \vertex [dot, fill=white, minimum size=3pt] (f3) at (3.5,0) {};
    \vertex [square dot] (f4) at (5,1.5) {};
    \vertex [square dot] (p1) at (0,-1.5) {};
    \vertex [square dot] (p2) at (5,-1.5) {};
    \diagram* {
      (f1) -- [plain] (f2) -- [boson] (f3) -- [plain] (f4),
      (f2) -- [plain] (p1),
      (f3) -- [plain] (p2),
    };
  \end{feynman}
\end{tikzpicture}}}
\label{Eq:Section_2.1)*The_four_possibilities_for_the_s_channel}
\end{align}
It is important to note that if the two vertices under consideration are distinct, we also must take into account that each diagram comes in two configurations that are derived by swapping the vertices.

\noindent\textbf{Fourth step:}
Then, we have to evaluate the amplitude associated with each diagram. First of all, we associate each external leg connected to an internal vertex with the so-called scalar bulk to boundary ($b$) propagator:
\begin{align}
    \vcenter{\hbox{\begin{tikzpicture}[ scale=0.5]
      \begin{feynman}
    \vertex [line width = 0.75 pt, dot,  minimum size=3pt] (f2) at (1.5,0) {};
    \vertex [square dot, minimum size=3pt] (f3) at (3.5,0) {};
    \diagram* {
      (f2) -- (f3),
    };
  \end{feynman}
\end{tikzpicture}}}\,=\,
G_{+b}(k,\tau,0) &= \hat{u}^{*}_{\zeta}(k,\tau)\hat{u}_{\zeta}(k,0)\,=\, \frac{H^{2}}{4\epsilon k^{3}M_{\mathrm{Pl}}^2}(1-ik\tau)e^{+ik\tau},\\
    \vcenter{\hbox{\begin{tikzpicture}[ line width = 0.75 pt, scale=0.5]
      \begin{feynman}
    \vertex [dot, fill=white, minimum size=3pt] (f2) at (1.5,0) {};
    \vertex [square dot, minimum size=3pt] (f3) at (3.5,0) {};
    \diagram* {
      (f2) -- (f3),
    };
  \end{feynman}
\end{tikzpicture}}}\,=\,G_{-b}(k,\tau,0)&= \hat{u}_{\zeta}(k,\tau)\hat{u}^{*}_{\zeta}(k,0) \,=\,\frac{H^{2}}{4\epsilon k^{3}M_{\mathrm{Pl}}^2}(1+ik\tau)e^{-ik\tau}.
\end{align}
\noindent\textbf{Fifth step:}
We associate to each internal graviton propagator the contribution for the mode functions
\begin{align}
    \vcenter{\hbox{\begin{tikzpicture}[line width = 0.75 pt, scale=0.7]
      \begin{feynman}
    \vertex [dot,  minimum size=3pt] (f2) at (1.5,0) {};
    \vertex [dot, minimum size=3pt] (f3) at (3.5,0) {};
    \diagram* {
      (f2) -- [boson] (f3),
    };
  \end{feynman}
\end{tikzpicture}}}\,=\,G_{++,s}(k,\tau_{1},\tau_{2}) &= \gamma_{s}(k,\tau_{1})\gamma^{*}_{s}(k,\tau_{2})\theta(\tau_{1}-\tau_{2}) + \gamma_{s}^{*}(k,\tau_{1})\gamma_{s}(k,\tau_{2})\theta(\tau_{2}-\tau_{1})\,,
    \nonumber\\
    \vcenter{\hbox{\begin{tikzpicture}[ line width = 0.75 pt, scale=0.7]
      \begin{feynman}
    \vertex [dot,  minimum size=3pt] (f2) at (1.5,0) {};
    \vertex [dot, fill=white, minimum size=3pt] (f3) at (3.5,0) {};
    \diagram* {
      (f2) -- [boson] (f3),
    };
  \end{feynman}
\end{tikzpicture}}}
\,=\,
    G_{+-,s}(k,\tau_{1},\tau_{2}) &= \gamma_{s}^{*}(k,\tau_{1})\gamma_{s}(k,\tau_{2})\,, 
    \nonumber\\
    \vcenter{\hbox{\begin{tikzpicture}[line width = 0.75 pt,  scale=0.7]
      \begin{feynman}
    \vertex [dot, fill=white, minimum size=3pt] (f2) at (1.5,0) {};
    \vertex [dot, minimum size=3pt] (f3) at (3.5,0) {};
    \diagram* {
      (f2) -- [boson] (f3),
    };
  \end{feynman}
\end{tikzpicture}}}
\,=\,
    G_{-+,s}(k,\tau_{1},\tau_{2}) &= \gamma_{s}(k,\tau_{1})\gamma_{s}^{*}(k,\tau_{2})\,, 
    \nonumber\\
    \vcenter{\hbox{\begin{tikzpicture}[line width = 0.75 pt, scale=0.7]
      \begin{feynman}
    \vertex [dot, fill=white, minimum size=3pt] (f2) at (1.5,0) {};
    \vertex [dot, fill=white, minimum size=3pt] (f3) at (3.5,0) {};
    \diagram* {
      (f2) -- [boson] (f3),
    };
  \end{feynman}
\end{tikzpicture}}}
\,=\,
    G_{--,s}(k,\tau_{1},\tau_{2}) &= \gamma_{s}^{*}(k,\tau_{1})\gamma_{s}(k,\tau_{2})\theta(\tau_{1}-\tau_{2}) + \gamma_{s}(k,\tau_{1})\gamma_{s}^{*}(k,\tau_{2})\theta(\tau_{2}-\tau_{1})\,.
    \label{Eq:G--,s}
\end{align}
We incorporate the helicity factor of the tensor mode functions within the vertex factor, as outlined in \cite{chen_schwinger-keldysh_2017}. It is important to emphasize that the helicity-dependent part of propagators remains identical for all four propagators.

\noindent\textbf{Sixth step:}
We associate to each vertex the respective vertex factor as described in \cite{chen_schwinger-keldysh_2017}. We want to stress that $\pm$ vertices are multiplied with $+$  and $-$ respectively. After completing this step, we integrate over $\tau_{1}$ and $\tau_{2}$ from $-\infty$, that is, the beginning of inflation, to $\tau_{f}$, that is, the end of inflation, which, in this instance, we set to zero. We recall that we associate each vertex with a conformal time variable, in this case $\tau_{1}$ and $\tau_{2}$.

\noindent\textbf{Seventh step:}
Multiply by $(2\pi)^3\delta^{(3)}(..)$, where $\delta^{(3)}(..)$ entails momentum conservation. Finally, sum over channels, vertices combination, and graviton helicities to get the final amplitude.
It should be noted that $(2\pi)^3\delta^{(3)}(..)$ will be excluded from the calculations; however, to accurately determine the signal-to-noise ratio of the parity-violating part of the trispectrum in section~\ref{Sec:The signal-to-noise ratio, GR-GR}, it must be reintroduced.

\subsection{The computation}
\label{Sec:the_computation}
In this section, we perform the original computation of the graviton-mediated trispectrum in chiral scalar-tensor theories of modified gravity. 
A priori one should consider all the scalar-scalar-tensor vertices (see section~\ref{Section:General framework for operating within the In-In formalism}) which can be used in a diagram such the one presented in eq.~\eqref{Eq:Parity_violating_trispectrum)The_graviton-mediated_trispectrum}. A detailed discussion of all the diagrams will be provided in \cite{workinprogress}, so we will not delve into them here. However, we provide an overview of the most important features. Regarding the graviton-mediated trispectrum we can have three different combinations for the scalar-scalar-tensor vertices
\begin{itemize}
    \item two parity-conserving, i.e.~GR vertices (see \cite{maldacena_non-gaussian_2003,ACQUAVIVA2003119}),
    \item two parity-violating vertices,
    \item a parity-conserving vertex and a parity-violating one.
\end{itemize}
For the first two scenarios, to produce a parity-violating graviton-mediated trispectrum we need the presence of parity-violating (chiral) graviton propagators, whereas for the last scenario, this requirement is not necessary. The trispectrum in the first scenario is very similar to the one computed with two GR vertices in the Chern-Simons theory of modified gravity, as presented in \cite{creque-sarbinowski_parity-violating_2023}, where it is noted that the signal is significantly suppressed. In the third case, as we will demonstrate in \cite{workinprogress}, generating a parity-violating trispectrum is not feasible without breaking the perturbative regime of the theory. Specifically, the amplitude of the free functions, i.e.~$a_A$, would need to be increased by three orders of magnitude beyond the maximum value allowed (see section~\ref{Sec:perturbativity_ratio} for the definition of the perturbativity ratio).
Thus, in this paper, we are going to consider the only case in which we can produce a detectable signal in a single-field slow-roll model of inflation within the chiral scalar-tensor theories of modified gravity.

The idea of this paper is to employ two parity-violating vertices. In particular, we use the PV vertices derived from the Lagrangian $\mathcal{L}_{\mathrm{PV}1}$ to take advantage of the free functions $a_A$ that, given eq.~\eqref{Eq:specific_parameters_configuration}, are not restricted by any constraints such as that in eq.~\eqref{Eq:chiral_condition}.
Moreover, we require chiral gravitons to generate a parity-violating graviton-mediated trispectrum while working with two parity-violating vertices. This is accomplished through the modifications introduced in the tensor equation of motion by $\mathcal{L}_{\mathrm{PV}2}$. We represent the situation in a diagrammatic way as shown in eq.~\eqref{Eq:PV_case_}.
\begin{align}
    \vcenter{\hbox{\begin{tikzpicture}[ scale=0.5]
      \begin{feynman}
    \vertex [label=left:\(\zeta(\mathbf{K}_{I}^{1})\), square dot] (f1) at (0,1.5) {};
    \vertex [label=left:\textcolor{pyred}{PV1}, dot, fill=pyred,minimum size=3pt] (f2) at (1.5,0) {};
    \vertex [dot, fill=pyred, label=right:\textcolor{pyred}{PV1}, minimum size=3pt] (f3) at (3.5,0) {};
    \vertex [square dot, label=right:\(\zeta(\mathbf{K}_{I}^{3})\)] (f4) at (5,1.5) {};
    \vertex [square dot, label=left:\(\zeta(\mathbf{K}_{I}^{2})\)] (p1) at (0,-1.5) {};
    \vertex [square dot, label=right:\(\zeta(\mathbf{K}_{I}^{4})\)] (p2) at (5,-1.5) {};
    \diagram* {
      (f1) -- [plain] (f2) -- [boson, edge label=\textcolor{orange}{\(\gamma_s(\mathbf{k}_{I})\)}] (f3) -- [plain] (f4),
      (f2) -- [plain] (p1),
      (f3) -- [plain] (p2),
    };
  \end{feynman}
\end{tikzpicture}}}
\label{Eq:PV_case_}
\end{align}

As we plan to exploit the free functions of $\mathcal{L}_{\mathrm{PV}1}$, i.e.~the $a_A$, to enhance the signal, we assert that it is  fully reasonable to disregard all the scalar-scalar-tensor vertices of $\mathcal{L}_{\mathrm{PV}2}$, due to the constraints imposed on the free functions with the chirality condition in eq.~\eqref{Eq:chiral_condition}.

Now, we need to expand $\mathcal{L}_{\mathrm{PV}1}$ to write down all the $\zeta\zeta\gamma$ vertices. However, we refrain from performing the full expansion. Instead, we perform the computation using a single type of vertex to demonstrate the detectability of the signal. We do not expect that the signal's order of magnitude will largely change when all vertices are taken into account and, including them is beyond the purpose of this paper.
Expanding $\mathcal{L}_{\mathrm{PV}1}$ to find couplings of the form $\varepsilon^{inj}\gamma_{mi}\zeta_{,n}\zeta^{,m}{}_{j}$ results in the $\zeta\zeta\gamma$ vertex 
\begin{align}
   \mathcal{L}_{\mathrm{PV}1}^{\zeta\zeta\gamma} =  8\epsilon\left(\frac{H}{M_{\mathrm{Pl}}}\right)^2\int \mathrm{d}\tau \mathrm{d}^{3}x\,a_1\varepsilon^{inj}\frac{\gamma_{mi}}{\tau}\zeta_{,n}\zeta^{,m}{}_{j}\,,
   \label{Eq:vertex_LPV1}
\end{align}
where $\mathcal{L}_{\mathrm{PV}1}^{\zeta\zeta\gamma}$ represents the part of the $\mathcal{L}_{\mathrm{PV}1}$ Lagrangian associated with the particular vertex form chosen.
Following the same approach of \cite{bartolo_tensor_2021}, we select a coupling form given by
\begin{align}
    a_1 = A \left(\frac{\tau_\ast}{\tau}\right)\,,
    \label{Eq:choice_a_1}
\end{align} 
where $A$ is real, and $\tau_\ast =14\,\mathrm{Gpc}$  is the time when the scale corresponding to present observable Universe crosses the horizon during inflation. Although $A$ is arbitrary, it is crucial to ensure that the perturbative treatment of the theory remains intact. In appendix \ref{Sec:perturbativity_ratio} we show that the maximum value allowed is 
\begin{align}
    A = \frac{1}{8\left(\frac{H}{M_\mathrm{Pl}}\right)^5(\tau_* k_\mathrm{max})}\,,
    \label{Eq:A}
\end{align}
where $k_\mathrm{max} = 0.3\,h/\mathrm{Mpc}$ is the maximum momentum that can be observed within the volume in which we are interested to evaluate the signal-to-noise ratio in section~\ref{Sec:The signal-to-noise ratio, GR-GR}.
This kind of coupling can arise through a dilaton-like coupling \cite{Shiraishi_2011} (see appendix \ref{Section:dilaton_like_coupling}). 
The choice \eqref{Eq:choice_a_1} for the free function $a_1$ may seem unconventional; however, dimensional analysis suggests that it will not alter the overall estimate of the signal-to-noise ratio of the parity-violating part of the trispectrum in section~\ref{Sec:The signal-to-noise ratio, GR-GR} with respect to the case in which we take the coupling function $a_1$ as constant. This choice is made to simplify the time integrals in the \textit{In-In} formalism, ensuring consistency with the shapes calculated in \cite{creque-sarbinowski_parity-violating_2023,Seery_2009}.

Therefore, we can determine the graviton-mediated trispectrum with the chosen vertices as the sum of the $s$, $u$ and $t$ channels as 
\begin{align}
     \Re\langle\zeta(\mathbf{k}_{1})\zeta(\mathbf{k}_{2})\zeta(\mathbf{k}_{3})\zeta(\mathbf{k}_{4})\rangle &= \sum_{I}\Re\left[P_{R}(\mathbf{K}_{I})\right]\left[T_R(\mathbf{K}_{I})+T_L(\mathbf{K}_{I})\right]\,,
     \label{Eq:Real_part_trispectrum}\\
      \Im\langle\zeta(\mathbf{k}_{1})\zeta(\mathbf{k}_{2})\zeta(\mathbf{k}_{3})\zeta(\mathbf{k}_{4})\rangle &= \sum_{I}\Im\left[P_{R}(\mathbf{K}_{I})\right]\left[T_R(\mathbf{K}_{I})-T_L(\mathbf{K}_{I})\right]\,,
      \label{Eq:Immaginary_part_trispectrum}
\end{align}
where 
$\langle\zeta(\mathbf{k}_{1})\zeta(\mathbf{k}_{2})\zeta(\mathbf{k}_{3})\zeta(\mathbf{k}_{4})\rangle$ represents the global trispectrum for this configuration. By global, we imply the sum of the three distinct channels, which results in the momentum dependence being $(\mathbf{k}_{1},\mathbf{k}_{2},\mathbf{k}_{3},\mathbf{k}_{4})$. On the right-hand side of eq.~\eqref{Eq:Real_part_trispectrum} and eq.~\eqref{Eq:Immaginary_part_trispectrum}, we recall that 
\begin{align}
    \mathbf{K}_{I} = (\mathbf{K}_{I}^{1},\mathbf{K}_{I}^{2},\mathbf{K}_{I}^{3},\mathbf{K}_{I}^{4})\,,
\end{align}
denotes the momentum dependencies for the three different channels (see section~\ref{Section:General framework for operating within the In-In formalism}). Moreover, $\Re\langle..\rangle$ and $\Im\langle..\rangle$ 
are the real and imaginary part of the total trispectrum. We also have considered the lowest order in slow-roll and introduced the polarization portion of the diagram, i.e.~the time-independent part of the computation, 
\begin{align}
    P_{s}(\mathbf{K}_{I}) &= 
    \varepsilon^{abc}\left[\epsilon^{s}{}_{af}(\mathbf{K}_{I})\right]^{\ast}\left(K_{I}^{1}\right)_{b}\left(K_{I}^{2}\right)_{c}\left(K_{I}^{2}\right)^{f} \varepsilon^{ijl}\epsilon^{s}_{in}(\mathbf{K}_{I}) \left(K_{I}^{3}\right)_{j}\left(K_{I}^{4}\right)_{l}\left(K_{I}^{4}\right)^{n}\nonumber\\&+ \left[\left(\mathbf{K}_{I}^1\leftrightarrow\mathbf{K}_{I}^2\right),\left(\mathbf{K}_{I}^3\leftrightarrow\mathbf{K}_{I}^4\right)\right]\,.
\end{align}
In eqs.~\eqref{Eq:Real_part_trispectrum} and \eqref{Eq:Immaginary_part_trispectrum} we have also introduced
\begin{align}
    T_{s}(\mathbf{K}_{I}) &= -2\epsilon^2\left(\frac{H}{M_{\mathrm{Pl}}}\right)^{-6}\left(\frac{1}{k_\mathrm{max}}\right)^2\Re\left[\mathcal{J}^{(2)}_{s}(\mathbf{K}_{I})-\mathcal{J}^{(1)}_{s}(\mathbf{K}_{I}^{1}, \mathbf{K}_{I}^{2})\mathcal{J}^{(1)}_{s}{}^{\ast}(\mathbf{K}_{I}^{3}, \mathbf{K}_{I}^{4})\right]\,,
    \label{Eq:T}
\end{align}
where we have adopted the following notation for the time integrals $\mathcal{J}^{(2)}_{s}(\mathbf{K}_{I})$ and $\mathcal{J}^{(1)}_{s}(\mathbf{K}_{I}^{i}, \mathbf{K}_{I}^{j})$ 
\begin{align}
    \mathcal{J}^{(2)}_{s}(\mathbf{K}_{I}) &=  \int_{-\infty}^{0} \frac{\mathrm{d}\tau_{1}}{ \tau_{1}^{2}}\int_{-\infty}^{0}\frac{\mathrm{d}\tau_{2}}{ \tau_{2}^{2}} 
    \mathcal{G}_{+}(\mathbf{K}_{I}^{1}, \tau_{1})
    \mathcal{G}_{+}(\mathbf{K}_{I}^{2}, \tau_{1})
    \mathcal{G}_{+}(\mathbf{K}_{I}^{3}, \tau_{2})
    \mathcal{G}_{+}(\mathbf{K}_{I}^{4}, \tau_{4})\times\nonumber\\ &\times\gamma_s[\mathbf{k}_{I}, \max(\tau_1, \tau_2)] \gamma^{\ast}_{s}[\mathbf{k}_{I}, \min(\tau_1, \tau_2)]\,, 
    \label{Eq:J_2}
    \\
    \mathcal{J}^{(1)}_{s}(\mathbf{K}_{I}^{i}, \mathbf{K}_{I}^{j}) &= \int_{-\infty}^{0}\frac{\mathrm{d}\tau}{ \tau^{2}} \mathcal{G}_{+}(\mathbf{K}_{I}^{i}, \tau) \mathcal{G}_{+}(\mathbf{K}_{I}^{j}, \tau) \gamma_{s}^{\ast}(\mathbf{K}_{I}^{i}+\mathbf{K}_{I}^{j}, \tau)\,,
    \label{Eq:J_1}
\end{align}
where $s=L/R$ is the polarization index, $\gamma_s(k,\tau)$ are the tensor mode functions (for the explicit expression, see equations (3.24), (3.25), (3.26), (3.27), (3.28) and (3.29) of ref. \cite{bartolo_tensor_2021}), and the scalar vertex to bulk propagator reads (see section~\ref{Section:General framework for operating within the In-In formalism})
\begin{align}
    \mathcal{G}_{+}({k},\tau)= \frac{H^2}{4\epsilon k^3M_{\mathrm{Pl}}^2}(1-i k\tau)e^{+i k \tau}\,,
\end{align}
where $k \,=\, \lvert\mathbf{k}\rvert$. Next, we assess the contribution from the polarization part, which we recall is the time-independent part, and the time integrals. 
\subsubsection{Polarization sum}\label{Sec:appendix_B.1.1*)Polarization_sum}
We know that is possible to decompose the spin-2 polarization matrices into a product of spin-1 polarization vectors of the same helicity as \cite{Shiraishi_2011}
\begin{align}
    \epsilon^{s}_{ij}(\mathbf{k}) = \sqrt{2}e_{i}^{s}(\mathbf{k})e_{j}^{s}(\mathbf{k})\,.
\end{align}
Using the orthogonality and completeness of these vectors over the $2$D complex plane, we get
\begin{align}
    \sqrt{2}e_{i}^{s}(\mathbf{k})e_{j}^{s\ast}(\mathbf{k}) = \frac{1}{2}\left(\delta_{ij} -\hat{k}_{i}\hat{k}_{j}-i\lambda_s\varepsilon_{ijl}\hat{k}^l\right)\,,
\end{align}
where $\lambda_s=\pm 1$ for right- and left-handed gravitons.
Using these properties, we can rewrite the polarization factor as
\begin{align}
    P_{\gamma}(\mathbf{K}_{I})=\frac{1}{2}k_{I}^2\left(K^{2}_{I}\right)^2\left(K^{4}_{I}\right)^2\sin^{2}\theta^{4}_{I}\sin^{2}\theta^{2}_{I}\exp{2i\lambda_s\left(\phi_{I}^{4} - \phi_{I}^{2}\right)}+\left(\mathbf{K}_{I}^1\leftrightarrow\mathbf{K}_{I}^2,\mathbf{K}_{I}^3\leftrightarrow\mathbf{K}_{I}^4\right)\,,
\end{align}
where we have chosen to work in the reference frame in which $\hat{\mathbf{e}}_{z} = \hat{\mathbf{k}}_{I}$. The angles $\theta_{I}^{i}$ and $\phi^i_I$ denote the polar and azimuthal coordinates, respectively, of $\mathbf{K}_{I}^{i}$.

\subsubsection{Time integral}\label{Sec:appendix_A.3.2*)Time_integral}
\begin{figure}[t]  
    \centering
    \includegraphics[width=1\textwidth]{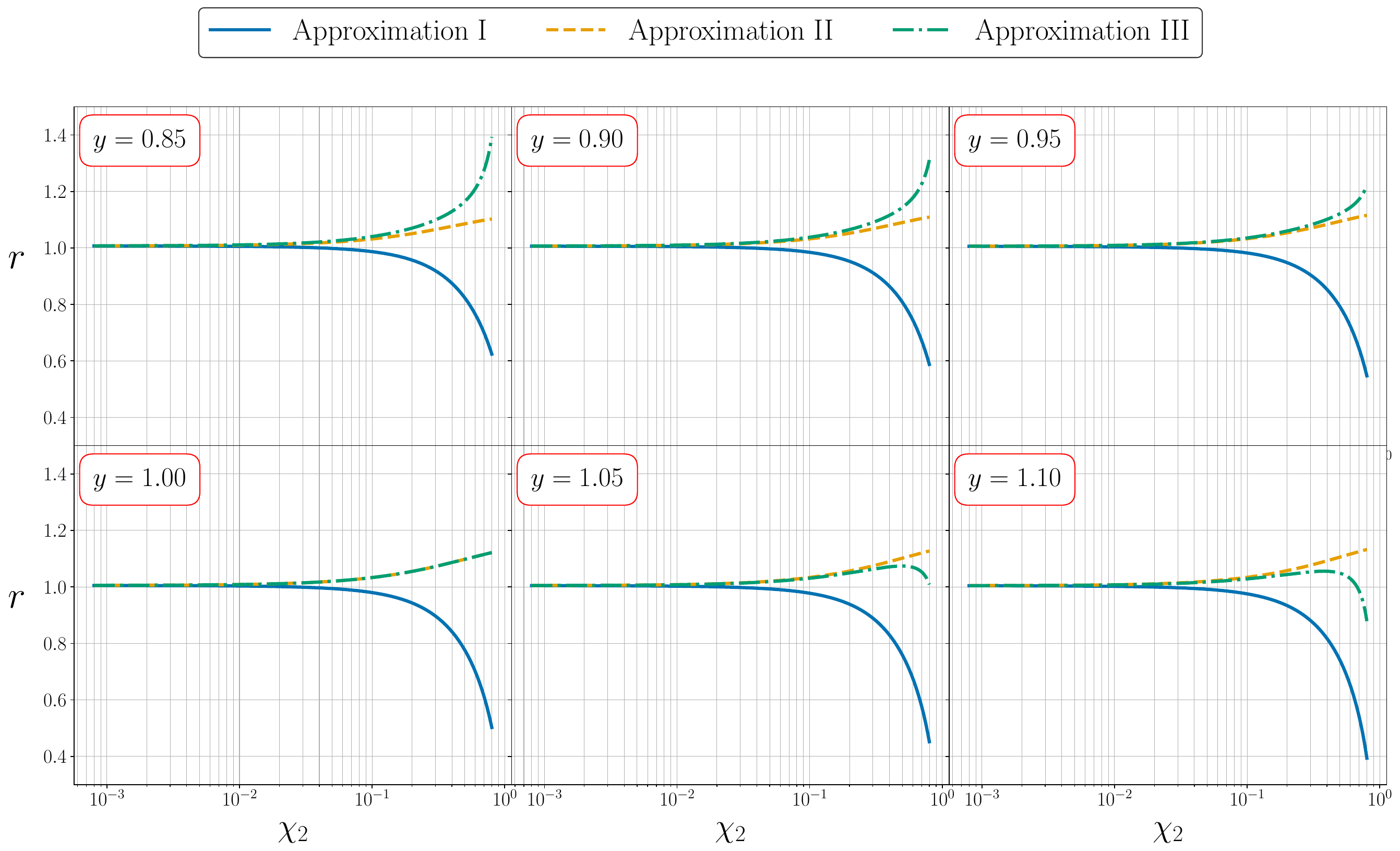}  
    \caption{We present the modulus of the ratio of the three approximate mode functions, i.e.~$\gamma_{s,\mathrm{I}}$ and $\gamma_{s,\mathrm{II}}$ and $\gamma_{s,\mathrm{III}}$, to the actual mode functions}
    \label{Fig:ratio_vs_chi}  
\end{figure}

We move forward to compute the time integral, i.e.~eqs.~\eqref{Eq:J_2} and \eqref{Eq:J_1}, using two different approximations for the tensor mode functions $\gamma_s$. Although there is an approximate analytic solution for tensor mode functions \cite{Qiao,Bartolo_2017}, evaluating the integrals using these expressions remains unfeasible. We approximate the tensor mode functions, using that we expect the actual mode functions to be the de Sitter ones (see \cite{Bartolo_2004}) with small corrective terms proportional to $\epsilon$, which are disregarded, and $\chi_2$.
\paragraph{Approximation I.} In the first case, we use the same approximation as in \cite{creque-sarbinowski_parity-violating_2023}. We take the tensor mode functions as the de Sitter ones scaled by a general complex number $C_s$
\begin{align}
    \gamma_{s,\,\mathrm{I}}(k,\tau)\approx C_s \frac{i H}{M_{\mathrm{Pl}}\sqrt{k^3}}(1+ik\tau)e^{-ik\tau}\,,
    \label{Eq:approximate_mode_functions_1}
\end{align}
where the subscript I indicates that these mode functions correspond to what we refer to as approximation I, $s$ denotes the polarization index, $\mathbf{k}$ stands for the three-momentum vector, and $k$ represents its modulus. To determine the value of $C_s$ we match the approximate solutions with the real ones in $\tau=0$.
Considering the expressions for the graviton propagator in eq.~\eqref{Eq:G--,s}, we note that $C_s$ appears only through its modulus. Hence, we can directly determine the modulus instead of $C_s$. We note that using the mode functions in eq.~\eqref{Eq:approximate_mode_functions_1}
results in the following tensor power spectrum
\begin{align}
    P_s = 2\,\lvert C_s\rvert^2 \left(\frac{H}{M_\mathrm{Pl}}\right)^2 = \frac{P_T^{\mathrm{GR}}}{2}\,\lvert C_s\rvert^2\,,
    \label{Eq:approximate_Power_spetrum}
\end{align}
where $P_T^{\mathrm{GR}}$ is the total tensor GR power spectrum at zero order in slow-roll. By matching eq.~\eqref{Eq:approximate_Power_spetrum} with eq.~\eqref{Eq:tensor_PS} we get
\begin{align}
    \lvert C_s\rvert^2 = \exp{\left(-\frac{\lambda_s\pi\chi_2}{16}\right)}\,.
    \label{Eq:square_modulus_C_s}
\end{align}
To discuss the validity of approximation I, as noted in \cite{creque-sarbinowski_parity-violating_2023}, we point out that the dominant contribution to integrals as those in eqs.~\eqref{Eq:J_2} and \eqref{Eq:J_1}, occurs near horizon crossing, i.e.~$-k\tau\approx1$. Thus, we can consider matching the approximate solutions at $\tau=0$ as a very rough estimate. Additionally, for sub-horizon values, the integrands are highly oscillatory, so their impact on the final result is negligible \cite{weinberg_quantum_2005}. Finally, in the plots of figure~\ref{Fig:ratio_vs_chi}, we present at different values of 
\begin{align}
    y \,=\, -k\tau \,=\,(0.85,\,0.9,\,0.95,\,1,\,1.05,\,1.1)\,,
\end{align}
the modulus of the ratio between the approximated mode function and the real one ($\gamma_{R}(\mathbf{k}, \tau)$) for right-handed gravitons, i.e.
\begin{align}
    r = \frac{\gamma_{R,\mathrm{I}}(\mathbf{k}, \tau)}{\gamma_{R}(\mathbf{k}, \tau)} \,,
    \label{Eq:modulus_ratio}
\end{align}
where the analytic expression for the tensor mode functions obtained in \cite{Qiao,bartolo_tensor_2021} reads as 

\begin{align}
     \gamma_{R}(\mathbf{k}, \tau)\,=\,\frac{u_{R}(\mathbf{k}, \tau, \chi_2)}{\sqrt{A^2_{T,R}}}\,.
\end{align}

In eq.~\eqref{Eq:approximate_mode_functions_1} we do not take into account the velocity birefringence effect introduced in the EoM by $\mathcal{L}_\mathrm{PV2}$ (see eq.~\eqref{PV2-quadratic}). Although this effect theoretically exists, we expect it to have a negligible impact because of the constraints on chirality in eq.~\eqref{Eq:chiral_condition}.
The calculations with left-handed gravitons are more complex, yet we do not expect different results. Our findings, which are consistent with those reported in \cite{creque-sarbinowski_parity-violating_2023}, suggest that the expected errors are of the order of $O(1)$ when $\chi_2 \leqslant 1.$ 
\paragraph{Approximation II.}
The second approximation, referred to as approximation II, which we employ, is 
\begin{align}
    \gamma_{s,\,\mathrm{II}}(k,\tau)\approx A_{s} \frac{i H}{M_{\mathrm{Pl}}\sqrt{k^3 (1+\lambda_s\chi_2k\tau)}}(1+ik\tau)e^{-ik\tau}\,,
    \label{Eq:approximate_mode_functions_2}
\end{align}
where $A_{s}$ is a complex number, $s$ is the polarization index, the subscript II is the analogous of what we have introduced in eq.~\eqref{Eq:approximate_mode_functions_1} and  
\begin{align}
    (1+\lambda_s\chi_2k\tau) = A^2_{T,s}\frac{2}{M_{\mathrm{Pl}}^2a^2}\,,
\end{align}
is the factor modifying the normalization factor $A^2_{T,s}$ (see eq.~\eqref{ATMPv2}) with respect to the GR case. In the same way that we have determined $\lvert C_s\rvert^2$ in eq.~\eqref{Eq:square_modulus_C_s} we get
\begin{align}
    \lvert A_s\rvert^2 = \exp{\left(-\frac{\lambda_s\pi\chi_2}{16}\right)}\,.
\end{align}
The difference with respect to approximation I is that we want to take into account the different time-dependent ``normalization'', i.e.~$A^2_{T,s}$, that shows up in these theories with respect to the GR case. We can use approximation II while working in the following regime
\begin{align}
   -\lambda_s\chi_2\tau k = \lambda_s\frac{k_\mathrm{phys}}{M_{\mathrm{PV}2}}\leq1\,,
\end{align}
which corresponds to the requirement of working within a well-behaved theory, as discussed in section~\ref{Sec:chiral_scalar_tensor)The EoM and the Power spectrum}. Concerning the validity of this approximation, the key concepts are those of approximation I. Moreover, the plots in figure~\ref{Fig:ratio_vs_chi} indicate that approximation II yields improved agreement compared to approximation I. In figure~\ref{Fig:ratio_vs_chi}, we plot the modulus of the ratio of the mode functions given in eq.~\eqref{Eq:approximate_mode_functions_2} compared to the actual one (see the solutions obtained in \cite{Qiao,bartolo_tensor_2021}), as done for approximation I  in eq.~\eqref{Eq:modulus_ratio}.
\paragraph{Approximation III.}
Nevertheless, using the mode functions as represented in eq.~\eqref{Eq:approximate_mode_functions_2} for computing the time integral is not feasible due to technical challenges. So we can make a further assumption by employing the following expression, which we name approximation III,
\begin{align}
    \gamma_{s,\,\mathrm{III}}(k,\tau)\approx A_{s} \frac{i H}{M_{\mathrm{Pl}}\sqrt{k^3 (1-\lambda_s\chi_2)}}(1+ik\tau)e^{-ik\tau}\,,
\end{align}
where we have evaluated $(1+\lambda_s\chi_2k\tau)$ at horizon crossing, as this is the region where the larger contributions from the time integrals are expected. The modulus of the ratio of this function to the actual one is shown in figure~\ref{Fig:ratio_vs_chi}. The plots suggest that approximation III is slightly more accurate percentage-wise than approximation I for $y\leq1$, and clearly better for $y\geq1$. For $y\leq1$, approximation III tends to overestimate the outcome, whereas approximation I underestimates it. Therefore, the true value is probably somewhere in the middle.
In the following, we present the results of the time integrals in eqs.~\eqref{Eq:J_2} and \eqref{Eq:J_1}, using approximations I and III.

We can employ approximations I and III to determine the time integrals. Before this, it is important to note that the tensor mode functions we are going to use, i.e.~$\gamma_{s,\,\mathrm{I}}$ in eq.~\eqref{Eq:approximate_mode_functions_1} and $\gamma_{s,\,\mathrm{III}}$ in eq.~\eqref{Eq:approximate_mode_functions_2}, correspond to the de Sitter ones scaled by a different amplitude for the left- and right-handed gravitons. Therefore, we can compute the time integrals using the de Sitter mode functions and then multiply the result by this amplitude. We get
\begin{align}
    T_{s}(\mathbf{K}_{I}) = f_j(\lambda_s,\chi_2)\left.T_{s}(\mathbf{K}_{I})\right|_{\chi_2=0}\,,
\end{align}
where $\left.T_{s}(\mathbf{K}_{I})\right|_{\chi_2=0}$  is the result of the time integrals evaluated with the de Sitter mode functions and $f_j(\lambda_s,\chi_2)$ is the different amplitude in the $j=\mathrm{I}\,/\,\mathrm{III}$ approximation
\begin{align}
    f_1(\lambda_s,\chi_2) \,=\, \exp{\left(-\frac{\lambda_s\pi\chi_2}{16}\right)}\,,\quad
    f_3(\lambda_s,\chi_2) \,=\, \frac{\exp{\left(-\frac{\lambda_s\pi\chi_2}{16}\right)}}{1-\lambda_s\chi_2}\,.
\end{align}
Note that $I$ denotes the particular channel we are currently dealing with, $I = s/t/u$, whereas $\mathrm{I}$ refers to approximation I.
Therefore, using \texttt{Mathematica} \cite{Mathematica} and adopting the $i\epsilon$ prescription in the asymptotic past \cite{maldacena_non-gaussian_2003}, we get (see \cite{Seery_2009} for the analytical derivation)
\begin{align}
    \left.T_s(\mathbf{K}_{I})\right|_{\chi_2=0} = &\,M\,\biggl\{ \frac{K_{I}^{1} + K_{I}^{2}}{[a_{34}^{I}]^{2}}\left[\frac{1}{2}(a_{34}^{I}+k_{I})([a_{34}^{I}]^{2}-2b_{34}^{I})+k_{I}^{2}(K_{I}^{3}+K_{I}^{4})\right] +(1,2) \leftrightarrow (3,4) \nonumber\\&
    + \frac{K_{I}^{1}K_{I}^{2}}{k_{t}}\left[\frac{b_{34}^{I}}{a_{34}^{I}}-k_{I}+\frac{k_{I}}{a_{12}^{I}}\left(K_{I}^{3}K_{I}^{4} -k_I\frac{b_{34}^{I}}{a_{34}^{I}}\right)\left(\frac{1}{k_{t}+a_{12}^{I}}\right)\right]+(1,2) \leftrightarrow (3,4)\nonumber\\&
    -\frac{k_{I}}{a_{12}^{I}a_{34}^{I}k_{t}}\left[b_{12}^{I}b_{34}^{I}+2k_{I}^{2}\left(\prod_{i=1}^{4}k_{i}\right)\left(\frac{1}{k_{t}^{2}}+\frac{1}{a_{12}^{I}a_{34}^{I}}+\frac{k_{I}}{k_{t}a_{12}^{I}a_{34}^{I}}\right)\right]\biggr\}\,,
    \label{Eq:appendix_B.1*)Time_integral_result}
\end{align}
with 
\begin{align}
    k_{t} &= \sum_{i=1}^{4}\lvert\mathbf{K}_{I}^{i}\rvert\,, \hspace{0.2cm}
    a_{ij}^{I} =\left[ \lvert\mathbf{K}_{I}^{i}\rvert + \lvert\mathbf{K}_{I}^{j}\rvert +k_{I}\right]\,, \hspace{0.2cm}
    b_{ij}^{I} = \left[(\lvert\mathbf{K}_{I}^{i}\rvert + \lvert\mathbf{K}_{I}^{j}\rvert)k_{I} +\lvert\mathbf{K}_{I}^{i}\rvert \lvert\mathbf{K}_{I}^{j}\rvert\right]\,, \hspace{0.2cm}\nonumber\\
        M &= 
        \left(\frac{H}{M_{\mathrm{Pl}}}\right)^{4}\left(\frac{1}{k_\mathrm{max}}\right)^2\left[2^7\epsilon^2 k_{I}^3\prod_{i=1}^{4}\left(K_{I}^{i}\right)^3\right]^{-1}\,.
\end{align}

We are now able to calculate the graviton-mediated trispectrum as
\begin{align}
    \Re\langle\zeta(\mathbf{k}_{1})\zeta(\mathbf{k}_{2})\zeta(\mathbf{k}_{3})\zeta(\mathbf{k}_{4})\rangle &= 2\sum_{I}\Re\left(P_{R}(\mathbf{K}_{I})\right)\left.T_R(\mathbf{K}_{I})\right|_{\chi_2=0}\,,\\
      \Im\langle\zeta(\mathbf{k}_{1})\zeta(\mathbf{k}_{2})\zeta(\mathbf{k}_{3})\zeta(\mathbf{k}_{4})\rangle &= [f_j(\lambda_R,\chi_2)-f_j(\lambda_L,\chi_2)]\sum_{I}\Im\left(P_{R}(\mathbf{K}_{I})\right)\left.T_R(\mathbf{K}_{I})\right|_{\chi_2=0}\,,
\end{align}
where we recall that the subscript $j=\mathrm{I}\,/\,\mathrm{III}$ depending on what approximation we are working with. At first order in $\chi_2$ we have
\begin{align}
    [f_\mathrm{I}(\lambda_R,\chi_2)-f_\mathrm{I}(\lambda_L,\chi_2)]&\simeq-\frac{\pi}{8}\chi_2\,,\\
    [f_\mathrm{III}(\lambda_R,\chi_2)-f_\mathrm{III}(\lambda_L,\chi_2)]&\simeq\left(2-\frac{\pi}{8}\right)\chi_2\,,
\end{align}
from which we understand that between the two different configurations we have an enhancement factor. So, from this analysis, we also understand that the choice of the approximating mode functions we use could have an impact.

\subsection{The signal-to-noise ratio of the trispectrum}\label{Sec:The signal-to-noise ratio, GR-GR}
To determine the signal-to-noise ratio (SNR) for this trispectrum, we adopt the approach outlined in \cite{creque-sarbinowski_parity-violating_2023}, which builds upon \cite{Smith:2015uia}. We consider a cosmic-variance-limited experiment that observes linear Fourier modes within the range 
\begin{align}
    k\in\left(k_{\mathrm{min}} = 0.003\,\frac{h}{\mathrm{Mpc}},\, k_{\mathrm{max}}=0.3\,\frac{h}{\mathrm{Mpc}}\right)\,,
\end{align}
in a comoving volume  
\begin{align}
    V\geq\left(\frac{2\pi}{k_{\mathrm{min}}}\right)^3\,.
\end{align}
Thus, the SNR of the parity-violating part, i.e.~the imaginary part of the trispectrum, becomes \cite{Smith:2015uia}
\begin{align}
    \left(\frac{S}{N}\right)^2 &= \frac{1}{24}\int_{k_{\mathrm{min}}}^{k_{\mathrm{max}}}\left[\prod_{i=1}^{4}\frac{\mathrm{d}^{3}\mathbf{k}_{i}}{(2\pi)^3}\right]\frac{\lvert\Im\langle\zeta(\mathbf{k}_{1})\zeta(\mathbf{k}_{2})\zeta(\mathbf{k}_{3})\zeta(\mathbf{k}_{4})\rangle\rvert^2}{P_{\zeta}(k_1)P_{\zeta}(k_2)P_{\zeta}(k_3)P_{\zeta}(k_4)}\,,
    \label{Eq: S/N}
\end{align}
where we recall that $P_{\zeta}(k)$ is the power spectrum of the gauge-invariant curvature perturbation $\zeta$ (see eq.~\eqref{Eq:scalar_power_spectrum}.
Since the integrand in eq.~\eqref{Eq: S/N} does not have any divergences, we evaluate it numerically. The results for the two approximations are
\begin{align}
    \left(\frac{S}{N}\right)_\mathrm{I} &\approx 0.76\left(\frac{\chi_2}{0.8}\right)\left(\frac{N_{\mathrm{modes}}}{10^6}\right)^{\frac{1}{2}}\left(\frac{0.3h/\mathrm{Mpc}}{k_\mathrm{max}}\right)^2
    \,,
\label{Eq:S/N_ratio_numerical_value_config_1}\\
    \left(\frac{S}{N}\right)_\mathrm{III} &\approx 3.12\left(\frac{\chi_2}{0.8}\right)\left(\frac{N_{\mathrm{modes}}}{10^6}\right)^{\frac{1}{2}}\left(\frac{0.3h/\mathrm{Mpc}}{k_\mathrm{max}}\right)^2\,,
    \label{Eq:S/N_ratio_numerical_value_config_2}
\end{align}
where we have used 
\begin{align}
    N_{\mathrm{modes}} = \frac{k^{3}_{\mathrm{max}}V}{6\pi^2}\,, \quad (2\pi)^3\delta^{3}(0) = V\,, \quad h=0.7\,.
\end{align}
The value of the parameter $A$ has been chosen by setting the perturbativity ratio to unity \cite{Leblond_2008}, as explained in the appendix \ref{Sec:perturbativity_ratio} (see eq.~\eqref{Eq:A} for the expression). We are using the fact that the only bound we have not to violate is that we want to keep the perturbative regime intact. To contextualize the value of the constant $A$, in appendix \ref{Sec:perturbativity_ratio}, we show that the ratio of the vertex used in the computation, i.e.~eq.~\eqref{Eq:vertex_LPV1}, to the GR one at horizon crossing results in a factor of $4 \times 10^4$.

In eqs.~\eqref{Eq:S/N_ratio_numerical_value_config_1} and \eqref{Eq:S/N_ratio_numerical_value_config_2}, we did not establish a detection benchmark at $3\sigma$. This is because the goal of the computation is to make an order-of-magnitude estimate of the result, and there are various sources of uncertainty affecting the exact value in question:
\begin{itemize}
    \item We have not taken into account all the vertices present in the theory.
    \item The upper limit for the free functions of the Lagrangian, i.e.~$a_1$, is derived using what we refer to as the perturbativity ratio in appendix \ref{Sec:perturbativity_ratio}, yet this too remains an order-of-magnitude estimate.
    \item As shown in the plots of figure~\ref{Fig:ratio_vs_chi}, depending on the approximation we make for the tensor mode functions, we can easily have a difference of a factor 4 as in eqs.~\eqref{Eq:S/N_ratio_numerical_value_config_1} and \eqref{Eq:S/N_ratio_numerical_value_config_2}. The plots seem to suggest that approximation III is better than approximation I. We have also not considered the velocity birefringence effect in the approximation of the tensor mode functions, which could also influence the outcome (even though it is expected to be small given the constraint in eq.~\eqref{Eq:chiral_condition}).
    \item The exact and precise value of the chirality can have an impact on the final result. It is important to note that we cannot saturate the chirality to one because the tensor modes encounter issues around horizon exit; that is, we will have 
\begin{align}
    -k\tau\chi_2\approx1\,,
\end{align}
which will result in a ill-defined normalization in the tensor mode functions; $A_{T,R}^2$ (see eq.~\eqref{ATMPv2}) will approach infinity.
\item As discussed in appendix \ref{Sec:perturbativity_ratio} it is possible to take into account the $k$-dependence in the maximum value of $A$ in a more precise way than what we have presented in eq.~\eqref{Eq:value_A}. As shown in the appendix, a simple guess as the one in eq.~\eqref{Eq:max_function} results in an enhancement factor of approximately 2.4 in eqs.~\eqref{Eq:S/N_ratio_numerical_value_config_1} and 
\eqref{Eq:S/N_ratio_numerical_value_config_2}. Considering this factor we would have
\begin{align}
    \left(\frac{S}{N}\right)_\mathrm{I}\approx 
    1.83\,,\quad\left(\frac{S}{N}\right)_\mathrm{III} \approx 7.48\,.
\end{align}

\end{itemize}
Considering all these caveats, we stress that the signal-to-noise ratio of the parity-violating part of the computed trispectrum is of order one and the perturbative regime of the theory is maintained. This implies that it is possible to produce a parity-violating graviton-mediated scalar trispectrum within chiral scalar-tensor theories of gravity without introducing changes in the action, i.e.~in eq.~\eqref{Section_1.1*):Global_action}, as the ones adopted, e.g., by \cite{creque-sarbinowski_parity-violating_2023}. This is the most important result of this paper since it provides a way within the single-field slow-roll inflationary model of chiral scalar-tensor theories of modified gravity to have a SNR of order one. All other possibilities that generate a detectable trispectrum will be discussed in \cite{workinprogress}. The main difference with respect to the Chern-Simons theory is that in CST theories of modified gravity we can take advantage of the free functions $a_A$ and $b_A$. This is not possible in CS because the free function, i.e.~the function $f(\phi)$ in eq.(2.6) of \cite{Bartolo_2017}, multiplying the CS interaction term, is bounded by a chiral condition as the one presented in eq.~\eqref{Eq:chiral_condition}. 

\section{Conclusions}\label{Sec:Conclusions}
In this paper, we have shown how to produce a detectable
parity-violating scalar trispectrum of the gauge-invariant curvature
perturbation $\zeta$ within single-field slow-roll models in a modified
gravity framework, without introducing modifications to the single-field
slow-roll setup. Parity violation arises in the tensor sector, thus
giving rise to a parity-violating graviton-mediated trispectrum. We achieved this goal within the chiral scalar-tensor theories of modified
gravity proposed in \cite{crisostomi_beyond_2018}, first analyzed for inflation in \cite{bartolo_tensor_2021}. 
This trispectrum is computed with two parity-violating vertices and chiral gravitons (section~\ref{Sec:Parity-violating-trispectrum}).  

We work in a specific parameter configuration for the functions $a_A$ and $b_A$ (see eq.~\eqref{Eq:specific_parameters_configuration}) that allows us to exploit the two features of the full theory relevant for enhancing the trispectrum: the free functions in the new scalar-scalar-tensor vertices and the chirality in the tensor EoM. The crucial step is to consider each effect as a modification of only one of the two Lagrangians. We use the vertices introduced by the first Lagrangian $\mathcal{L}_{\mathrm{PV}1}$ while the chirality in the tensor EoMs arises from the second, $\mathcal{L}_{\mathrm{PV}2}$. It is necessary to separate the two effects because, in an arbitrary parameter configuration, both Lagrangians modify the tensor equations of motion, which results in the free functions $a_A$ and $b_A$ being constrained by chiral conditions (eq.~\eqref{Eq:chiral_condition}). Therefore, we have chosen to work in a configuration where the first Lagrangian does not modify the tensor equations of motion. This ensures that the only constraint on the value of $a_A$ is the perturbative bound \cite{Leblond_2008}.

Then, we have presented an order-of-magnitude estimate of the signal-to-noise ratio of the parity-violating part of the computed trispectrum (section~\ref{Sec:The signal-to-noise ratio, GR-GR}), which is approximately of order one. 
Although the hints for parity violation in the galaxy 4PCF \cite{hou_measurement_2022,philcox2021detection} have been questioned, we expect that this signal could potentially serve as a source for these observations or that upper bounds on such parity violation in the 4PCF of galaxies can constrain these theories.

We point out that other scenarios within these theories can be investigated to generate a detectable signal. However, it is necessary to modify the setup we are working with, i.e.~the single-field slow-roll model of inflation in the chiral scalar-tensor theories of modified
gravity. These aspects will be addressed in a forthcoming paper \cite{workinprogress}. It would also be interesting to perform a data analysis testing the various predictions of these models (not only the four-point correlation function studied here, but corresponding bispectra, like the scalar-scalar-tensor one or the tensor-tensor-tensor bispectrum studied in~\cite{bartolo_tensor_2021}, using different observables by using a dedicated (e.g.~template fitting) analysis, which, however, goes beyond the goal of this paper.

\acknowledgments
We are grateful to the referee for their valuable comments and insights, which have helped improve the manuscript.
We thank Angelo Caravano, Eiichiro Komatsu, Zachary Slepian, and Matteo Braglia for valuable discussions. We also convey our gratitude towards the Institute for Fundamental Physics of the Universe (IFPU), which hosted the focus week ``Parity violation through CMB observations'',\footnote{\url{https://www.ifpu.it/focus-week-2024-05-27/}} where we enjoyed many useful discussions. 
TM acknowledges the FCT project with ref. number PTDC/FIS-AST/0054/2021.
NB acknowledges financial support from the INFN InDark initiative and from the COSMOS network (www.cosmosnet.it) through the ASI (Italian Space Agency) Grants 2016-24-H.0, 2016-24-H.1-2018 and 2020-9-HH.0.
NB acknowledges support by the MUR PRIN2022 Project “BROWSEPOL: Beyond standaRd mOdel With coSmic microwavE background POLarization”-2022EJNZ53 financed by the European Union - Next Generation EU.

\newpage
\appendix
\section{Polarization tensors}\label{Polarization tensors}
The polarization tensors in the chiral basis, i.e.~in the basis of $L$ and $R$ circular polarization states, are traceless and transverse tensors, defined in term of the + and $\times$ basis as
\begin{align}
    \epsilon_{ij}^{R}(\mathbf{k}) = \frac{\epsilon_{ij}^{+}(\mathbf{k})+i\epsilon_{ij}^{\times}(\mathbf{k})}{\sqrt{2}}\,, \quad\quad  \epsilon_{ij}^{L}(\mathbf{k}) = \frac{\epsilon_{ij}^{+}(\mathbf{k})-i\epsilon_{ij}^{\times}(\mathbf{k})}{\sqrt{2}}\,.
\end{align}
It is possible to show that the following relations hold \cite{alexander_birefringent_2005}
\begin{align}
    \left[\epsilon_{ij}^{L}\right]^{*}(\mathbf{k}) &= \epsilon_{ij}^{R}(\mathbf{k})\,,\nonumber \\
        \epsilon^{(s)}{}^{i}{}_{i}(\mathbf{k}) &= 0\,, \nonumber\\
        \epsilon_{ij}^{(s)}(\mathbf{k})k^{i} &=  0\,,\nonumber\\
        \epsilon_{ij}^{R}(-\mathbf{k}) &= \epsilon_{ij}^{L}(\mathbf{k})\,,\nonumber\\
    \epsilon^{ij}_{R}(\mathbf{k})\epsilon_{ij}^{L}(\mathbf{k}) &= 2\,, \nonumber\\
    \epsilon^{ij}_{R}(\mathbf{k})\epsilon_{ij}^{R}(\mathbf{k})&= 0\,,\nonumber\\
    \varepsilon^{ijf}k_{j}\epsilon^{(s)}_{f}{}^{l}(\mathbf{k}) &= -i\lambda_{s}k\epsilon^{(s)}{}^{li}(\mathbf{k})\,,
    \label{Polar}
\end{align}
where $\varepsilon^{ijf}$ is the three-dimensional Levi-Civita symbol, $k=\lvert\mathbf{k}\rvert$ and we have used Einstein's convention for contracted indices.

\section{The dilaton-like coupling form}
\label{Section:dilaton_like_coupling}
Before going into the details, we notice that our choice of the coupling function $a_1$ in eq.~\eqref{Eq:choice_a_1} should be viewed as an illustrative ad-hoc example. This draws inspiration from dilaton models but does not correspond to any presently established ultraviolet (UV) completion. The purpose here is to provide an example rather than a fully consistent model rooted in a known fundamental theory. We stress once again that this choice was made to simplify the time integrals in the \textit{In-In} formalism. A priori, one could have treated the coupling $a_1$ as a constant and performed the computation accordingly.

Next, we show how to derive the time dependence in eq.~\eqref{Eq:choice_a_1} by assuming a dilaton-like coupling.
In single-field slow-roll models of inflation, the EoM for the background inflaton field in conformal time reads \cite{Bartolo_2004}
\begin{align}
    \phi^{'}_{0}\approx \pm\frac{\sqrt{2\epsilon}M_{\mathrm{Pl}}}{\tau}\,,
\end{align}
where $'\equiv{\partial}/{\partial\tau}$, the subscript 0 recalls that we are working at the background level and the $+$ and $-$ sign refers respectively to ${\partial V(\phi)}/{\partial\phi}>0$ and ${\partial V(\phi)}/{\partial\phi}<0$. This equation can be integrated over time to give
\begin{align}
    \phi_{0} = \bar{\phi_{0}} \pm \sqrt{2\epsilon}M_{\mathrm{Pl}}\ln{\left(\frac{\tau}{\bar{\tau}}\right)}\,,
    \label{Eq:EoM_inflaton_solution}
\end{align}
where $\bar{\tau}$ is a fixed instant in time. Since the free functions $ a_s$ depend on the inflaton field and its kinetic term (section~\ref{Sec:Chiral_scalar-tensor_theories_of_gravity}), we can select for the coupling $a_1$ this specific functional dependence
\begin{align}
    a_1 = A\exp{\left[\pm\frac{\phi-\bar{\phi}}{M}\right]} = A\exp{\left[\pm\frac{\phi_{0}-\bar{\phi}_0}{M}\right]}\,,
\end{align}
where $M$ is some arbitrary energy scale, $A$ is a constant, and in the second equality we have used that we are working in the comoving gauge (section~\ref{Sec:chiral_scalar_tensor)The actions_in_the_ADM_framework}). Now, by setting $M=\sqrt{2\epsilon}M_{\mathrm{Pl}}$ and $\bar{\tau}=\tau_{*}$, we can use eq.~\eqref{Eq:EoM_inflaton_solution} to get the desired coupling 
\begin{align}
    a_1  = A\exp{\left[\pm\frac{\phi_{0}-\bar{\phi}_0}{M}\right]} = A\left(\frac{\tau_*}{\tau}\right)\,.
\end{align}

\section{Perturbativity ratio}
\label{Sec:perturbativity_ratio}

In this appendix, we provide a concise overview of the concepts discussed in \cite{Leblond_2008} and use it to bound the parameter value $A$.

The method builds on the idea that to work within the perturbative regime of the theory, it is necessary to enforce a set of conditions on the action to ensure that the perturbative approach remains valid. Thus, we can think of writing the inflationary action, as the one in eq.~\eqref{Section_1.1*):Global_action}, in powers of the fluctuations
\begin{align}
    \mathcal{S} = \mathcal{S}_{0} + \mathcal{S}_2+\mathcal{S}_3 + .. \,\,,
\end{align}
where $\mathcal{S}_0$ is the background action, which is the homogeneous part, i.e.~the purely time-dependent part, describing the background dynamics of the model, which we recall, in this case, is identical to the one described in the standard model of single-field slow-roll inflation \cite{Bartolo_2004}. The $\mathcal{S}_1$ is zero if we substitute, as is normally done, the background fields with their solutions of the equations of motion. The term $\mathcal{S}_2$ represents the quadratic component which allows us to derive the EoMs for the perturbations, i.e.~$\zeta$ and $\gamma_{ij}$, while terms of higher order denote interaction terms.
In order to ensure that the perturbative treatment remains valid \cite{Leblond_2008} we have to impose at horizon crossing that
\begin{align}
    \mathcal{S}_0&> \mathcal{S}_2\,,\nonumber\\
    \mathcal{S}_i&\geq\mathcal{S}_j\,, \,\, i< j\,\in\mathbb{N}\,.  \label{Eq:Perturbativity_conditions}
\end{align}
Evaluating at horizon crossing basically consists in taking the Lagrangian and estimating the fields and the derivatives in the following way

\begin{align}
    \frac{\partial}{\partial\tau}\sim aH\,,\quad\frac{\partial}{\partial x}\sim aH\,,\quad\zeta\sim\frac{H}{M_\mathrm{Pl}\sqrt{\epsilon}}\,, \quad \gamma_s\sim \frac{H}{M_\mathrm{Pl}}\,,
\end{align}
where $\epsilon$ is the first slow-roll parameter.  Please note that we are neglecting modifications to the tensor speed, i.e.~we consider one in this case. We can do this because the correction we introduce is a function of $\chi_2$ and is the same for left- and right-handed gravitons. Therefore, it cancels out when calculating something similar to \eqref{Eq:Immaginary_part_trispectrum}. Now, we can apply the conditions of eq.~\eqref{Eq:Perturbativity_conditions} to bound the free parameter $A$ in the scalar-scalar-tensor vertex, that we have used in computing the trispectrum, which we recall is
\begin{align}
   \mathcal{L}_{\mathrm{PV}1}^{\zeta\zeta\gamma} =  8\epsilon\left(\frac{H}{M_{\mathrm{Pl}}}\right)^2A\left(\frac
   {\tau_*}{\tau}\right)\varepsilon^{inj}\frac{\gamma_{mi}}{\tau}\zeta_{,n}\zeta^{,m}{}_{j}\,.
\end{align}
Thus, we can define what we referred to in the main text as the perturbativity ratio as
\begin{align}
    r_p \equiv \left.\frac{\mathcal{L}_{\mathrm{PV}1}^{\zeta\zeta\gamma}}{\mathcal{L}_{\zeta\zeta}}\right|_{k = aH}\approx 8\left(\frac{H}{M_{\mathrm{Pl}}}\right)^5\left(\tau_*k\right)A\,,
    \label{Eq:perturbativity_ratio}
\end{align}
where $\mathcal{L}_{\zeta\zeta}$ is the Lagrangian at second order in scalar perturbations.
Now, taking as\footnote{See section \ref{Sec:The signal-to-noise ratio, GR-GR} for the definition of $k_{\mathrm{max}}$.} 
\begin{align}
    k=k_{\mathrm{max}} = 0.3\,\frac{h}{\mathrm{Mpc}}\,,
\end{align}
and setting $r_p=1$ we obtain an estimate of the maximum value that the free parameter $A$ can take
\begin{align}
    A = \frac{1}{8\left(\frac{H}{M_{\mathrm{Pl}}}\right)^5\left(\tau_*k_\mathrm{max}\right)}\,.
    \label{Eq:value_A}
\end{align}
To better contextualize the value of the constant $A$, we provide a comparison between the vertex used in our case and the scalar-scalar-tensor vertex calculated in GR, which is \cite{maldacena_non-gaussian_2003,ACQUAVIVA2003119}
\begin{align}
    \mathcal{L}_{\mathrm{GR}}^{\zeta\zeta\gamma} =  \epsilon\left(\frac{M_{\mathrm{Pl}}}{H}\right)^2\int \frac{\mathrm{d}\tau \mathrm{d}^{3}x}{\tau^2}\gamma_{mi}\zeta^{,i}\zeta^{,m}\,.
    \label{Eq:scalar_scalar_tensor_GR}
\end{align}
The comparison of the two vertices at horizon crossing reads
\begin{align}
    r = \left.\frac{\mathcal{L}_{\mathrm{PV}1}^{\zeta\zeta\gamma}}{\mathcal{L}_{\mathrm{GR}}^{\zeta\zeta\gamma}}\right|_{k=aH}\approx 8A\left(\frac{H}{M_{\mathrm{Pl}}}
\right)^4 (\tau_{*}k_\mathrm{max})\approx \left(\frac{M_{\mathrm{Pl}}}{H}
\right)\approx4\times10^{4}\,,
\label{Eq:ratio_vertices}
\end{align}
where $M_\mathrm{Pl}$ is the reduced Planck mass and we have used the current upper-bound on the inflationary Hubble scale $H \approx \SI{5e13}{\giga\electronvolt}$ \cite{Planck_2020_X}.
To conclude this section, we would like to make a remark regarding the possibility of accounting for the $k$-dependence in eq.~\eqref{Eq:perturbativity_ratio} in a more accurate way than simply setting $k = k_\mathrm{max}$. In fact, this can be done. We can consider using something like
\begin{align}
    k = \mathrm{Max}\,[K_I^1,K_I^2,k_I]\,,
    \label{Eq:max_function}
\end{align}
where we recall that $K_I^1$ and $K_I^2$ are the modulus of the two scalar momenta entering the vertex, while $k_I$ is the one of the graviton (see section~\ref{Sec:Parity-violating-trispectrum}).\footnote{In the GM trispectrum we have two vertices: one involving the three momenta $\mathbf{K}_I^1$, $\mathbf{K}_I^2$ and $\mathbf{k}_I$ and the other involving $\mathbf{K}_I^3$, $\mathbf{K}_I^4$ and $\mathbf{k}_I$. Therefore, the argument of $\mathrm{Max[..]}$ must be adjusted accordingly.}
We have calculated the signal-to-noise ratio in eqs.~\eqref{Eq:S/N_ratio_numerical_value_config_1} and \eqref{Eq:S/N_ratio_numerical_value_config_2} with this additional assumption, and we do not observe a significant difference in the result. An enhancement factor of 2.4 is obtained, but this does not have a substantial impact on the final outcome.

\bibliography{main.bib}

\end{document}